\documentclass[11pt,a4paper]{article}
\pdfoutput=1
\usepackage[utf8]{inputenc}
\usepackage{jheppub}
\usepackage{microtype}
\usepackage{braket}
\usepackage{amsthm,amsbsy,amsfonts,mathrsfs,enumerate,float,wrapfig,amsmath,mathtools,graphicx,framed,tcolorbox,amssymb}
\usepackage{bbm}
\usepackage{tikz}
\usepackage{todonotes}
\usepackage{subfigure}
\usepackage{graphbox}
\usepackage{tkz-euclide}

\newcommand{\lt}{\left}
\newcommand{\rt}{\right}
\newcommand{\frakA}{\ensuremath{\mathfrak{A}}}

\newcommand{\frakR}{\ensuremath{\mathfrak{R}}}

\usepackage{tikz}
\usetikzlibrary{patterns}

\tikzset{
pattern size/.store in=\mcSize, 
pattern size = 5pt,
pattern thickness/.store in=\mcThickness, 
pattern thickness = 0.3pt,
pattern radius/.store in=\mcRadius, 
pattern radius = 1pt}
\makeatletter
\pgfutil@ifundefined{pgf@pattern@name@_32nusnocv}{
\pgfdeclarepatternformonly[\mcThickness,\mcSize]{_32nusnocv}
{\pgfqpoint{0pt}{0pt}}
{\pgfpoint{\mcSize+\mcThickness}{\mcSize+\mcThickness}}
{\pgfpoint{\mcSize}{\mcSize}}
{
\pgfsetcolor{\tikz@pattern@color}
\pgfsetlinewidth{\mcThickness}
\pgfpathmoveto{\pgfqpoint{0pt}{0pt}}
\pgfpathlineto{\pgfpoint{\mcSize+\mcThickness}{\mcSize+\mcThickness}}
\pgfusepath{stroke}
}}
\makeatother

\tikzset{
pattern size/.store in=\mcSize, 
pattern size = 5pt,
pattern thickness/.store in=\mcThickness, 
pattern thickness = 0.3pt,
pattern radius/.store in=\mcRadius, 
pattern radius = 1pt}
\makeatletter
\pgfutil@ifundefined{pgf@pattern@name@_j87jngvy3}{
\pgfdeclarepatternformonly[\mcThickness,\mcSize]{_j87jngvy3}
{\pgfqpoint{0pt}{0pt}}
{\pgfpoint{\mcSize+\mcThickness}{\mcSize+\mcThickness}}
{\pgfpoint{\mcSize}{\mcSize}}
{
\pgfsetcolor{\tikz@pattern@color}
\pgfsetlinewidth{\mcThickness}
\pgfpathmoveto{\pgfqpoint{0pt}{0pt}}
\pgfpathlineto{\pgfpoint{\mcSize+\mcThickness}{\mcSize+\mcThickness}}
\pgfusepath{stroke}
}}
\makeatother

\tikzset{
pattern size/.store in=\mcSize, 
pattern size = 5pt,
pattern thickness/.store in=\mcThickness, 
pattern thickness = 0.3pt,
pattern radius/.store in=\mcRadius, 
pattern radius = 1pt}
\makeatletter
\pgfutil@ifundefined{pgf@pattern@name@_4fqllij1c}{
\pgfdeclarepatternformonly[\mcThickness,\mcSize]{_4fqllij1c}
{\pgfqpoint{0pt}{0pt}}
{\pgfpoint{\mcSize+\mcThickness}{\mcSize+\mcThickness}}
{\pgfpoint{\mcSize}{\mcSize}}
{
\pgfsetcolor{\tikz@pattern@color}
\pgfsetlinewidth{\mcThickness}
\pgfpathmoveto{\pgfqpoint{0pt}{0pt}}
\pgfpathlineto{\pgfpoint{\mcSize+\mcThickness}{\mcSize+\mcThickness}}
\pgfusepath{stroke}
}}
\makeatother

\tikzset{
pattern size/.store in=\mcSize, 
pattern size = 5pt,
pattern thickness/.store in=\mcThickness, 
pattern thickness = 0.3pt,
pattern radius/.store in=\mcRadius, 
pattern radius = 1pt}
\makeatletter
\pgfutil@ifundefined{pgf@pattern@name@_0hz0ptd8u}{
\pgfdeclarepatternformonly[\mcThickness,\mcSize]{_0hz0ptd8u}
{\pgfqpoint{0pt}{0pt}}
{\pgfpoint{\mcSize+\mcThickness}{\mcSize+\mcThickness}}
{\pgfpoint{\mcSize}{\mcSize}}
{
\pgfsetcolor{\tikz@pattern@color}
\pgfsetlinewidth{\mcThickness}
\pgfpathmoveto{\pgfqpoint{0pt}{0pt}}
\pgfpathlineto{\pgfpoint{\mcSize+\mcThickness}{\mcSize+\mcThickness}}
\pgfusepath{stroke}
}}
\makeatother

\tikzset{
pattern size/.store in=\mcSize, 
pattern size = 5pt,
pattern thickness/.store in=\mcThickness, 
pattern thickness = 0.3pt,
pattern radius/.store in=\mcRadius, 
pattern radius = 1pt}
\makeatletter
\pgfutil@ifundefined{pgf@pattern@name@_ngc26pfiu}{
\pgfdeclarepatternformonly[\mcThickness,\mcSize]{_ngc26pfiu}
{\pgfqpoint{0pt}{0pt}}
{\pgfpoint{\mcSize+\mcThickness}{\mcSize+\mcThickness}}
{\pgfpoint{\mcSize}{\mcSize}}
{
\pgfsetcolor{\tikz@pattern@color}
\pgfsetlinewidth{\mcThickness}
\pgfpathmoveto{\pgfqpoint{0pt}{0pt}}
\pgfpathlineto{\pgfpoint{\mcSize+\mcThickness}{\mcSize+\mcThickness}}
\pgfusepath{stroke}
}}
\makeatother

\tikzset{
pattern size/.store in=\mcSize, 
pattern size = 5pt,
pattern thickness/.store in=\mcThickness, 
pattern thickness = 0.3pt,
pattern radius/.store in=\mcRadius, 
pattern radius = 1pt}
\makeatletter
\pgfutil@ifundefined{pgf@pattern@name@_u58tsbgr4}{
\pgfdeclarepatternformonly[\mcThickness,\mcSize]{_u58tsbgr4}
{\pgfqpoint{0pt}{0pt}}
{\pgfpoint{\mcSize+\mcThickness}{\mcSize+\mcThickness}}
{\pgfpoint{\mcSize}{\mcSize}}
{
\pgfsetcolor{\tikz@pattern@color}
\pgfsetlinewidth{\mcThickness}
\pgfpathmoveto{\pgfqpoint{0pt}{0pt}}
\pgfpathlineto{\pgfpoint{\mcSize+\mcThickness}{\mcSize+\mcThickness}}
\pgfusepath{stroke}
}}
\makeatother
\tikzset{every picture/.style={line width=0.75pt}} 

\usetikzlibrary{positioning}
\usetikzlibrary{arrows, arrows.meta}
\usetikzlibrary{decorations, decorations.markings, decorations.pathmorphing}

\title{ O-vertex, O7$^+$-plane, and Topological Vertex}

\author[a]{Sung-Soo Kim,}
\author[b]{Xiaobin Li,}
\author[c]{Futoshi Yagi,}
\author[d]{Rui-Dong Zhu}
 \affiliation[a]{School of Physics, University of Electronic Science and Technology of China, \\
 No.2006 Xiyuan Ave, West Hi-Tech Zone, Chengdu, Sichuan 611731, China}
 \affiliation[b]{School of Mathematics, Southwest Jiaotong University,\\
 West zone, High-tech district, Chengdu, Sichuan 611756, China}
 \affiliation[c]{School of Science, Huzhou University,
 759 Erhuan East Road, Huzhou, Zhejiang, 313000, China}
 \affiliation[d]{Institute for Advanced Study \& School of Physical Science and Technology, Soochow University, Suzhou 215006, China}

\emailAdd{sungsoo.kim@uestc.edu.cn}
\emailAdd{lixiaobin@home.swjtu.edu.cn}
\emailAdd{60714@zjhu.edu.cn}
\emailAdd{rdzhu@suda.edu.cn}

\abstract{We revisit the instanton partition function for 5d $\mathcal{N}=1$ SO($N$) gauge theories compactified on S$^1$, computed from the topological vertex formalism with the O-vertex based on a 5-brane web diagram with an O5-plane. We introduce an identity that enables us to rewrite the unrefined partition function into a new expression in terms of the Nekrasov factors summed over Young diagrams, which can be interpreted as the freezing of an O7-plane. Based on this, we propose topological vertex formalism with an O7$^+$-plane.}

\begin{document}

\maketitle

\section{Introduction}

The topological vertex formalism \cite{Aganagic:2003db, Iqbal:2007ii, Awata:2008ed} is a power tool to study five-dimensional (5d) $\mathcal{N}=1$ supersymmetric gauge theory compactified on an S$^1$. When the corresponding toric Calabi-Yau threefolds are given, we can compute the topological string partition functions, which are identified as Nekrasov instanton partition functions \cite{Nekrasov:2002qd, Nekrasov:2003rj}.

From the perspective of Type IIB string theory, a $\mathcal{N}=1$ supersymmetric gauge theory can be described by a 5-brane web diagram \cite{Aharony:1997ju, Aharony:1997bh,Benini:2009gi}. Thanks to the equivalence \cite{Leung:1997tw} between the toric diagrams and the 5-brane web diagrams, one can apply the topological vertex formalism to compute the partition function for the theory on the Omega background. Recent development \cite{Kim:2017jqn, Bourgine:2017rik,Kim:2022dbr} has demonstrated that the topological vertex is applicable to 5-brane webs with an O5$^-$- or ON$^-$-plane. This technique has been used for computing the partition function for various theories, including the following gauge groups: Sp($N$)  \cite{Kim:2017jqn, Li:2021rqr}, SO($N$)\cite{Kim:2021cua}, and even G$_2$ \cite{Hayashi:2018bkd}, and also D-type quiver gauge theories \cite{Bourgine:2017rik, Kim:2022dbr}. Topological vertex for six-dimensional theories with/without $\mathbb{Z}_2$ twist is also studied in \cite{Kim:2021cua,Kim:2020npz}.

A noticeable generalization of the topological vertex for a 5-brane web with an O5-plane was introduced by a new vertex called ``O-vertex'' \cite{Hayashi:2020hhb} which can be understood as extending O5$^+$-plane to infinity. This makes it systemized to compute the Nekrasov instanton partition function for 5d $\mathcal{N}=1$ SO($2N$) gauge theories.

Originally, the SO($2N$) partition functions are given \cite{Nekrasov:2004vw} in the integral form based on ADHM construction. Unlike the U($N$) case, the expression for it in terms of the sum over the Young diagrams had not been known for a long time. With the help of the O-vertex, an expression in terms of the sum over Young diagrams for the unrefined case has been successfully obtained \cite{Hayashi:2020hhb, Nawata:2021dlk}, which agrees with the results obtained by computing the JK-residues of the integral form.

The SO($2N$) gauge theories can also be realized with an O7$^+$-plane \cite{Bergman:2015dpa} instead of an O5$^{\pm}$-plane. While an O5-plane appears as a line on the 5-brane web and is associated with a plane reflection, an O7-plane is a point or a dot on the 5-brane web giving rise to a $\pi$-rotation. Although these two 5-brane web diagrams look similar, there are significant differences regarding the intersection with 5-branes and the orientifold plane. For the O5-plane case, one can assign the O-vertex at the intersection between 5-branes and the O5-plane. On the other hand, for the O7-plane case, the 5-branes do not intersect with the O7-plane in general, as the O7-plane has the corresponding branch cut.

Despite such differences, a method to compute the SO($2N$) partition functions based on the 5-brane web diagram with an O7-plane based on the topological vertex formalism would be highly desirable as it would provide an intuitive understanding of the contributions to the partition function based on a 5-brane web involving an O7-plane (including hypermultiplets in antisymmetric or symmetric representations) and may also provide an alternative and systematic tool for computing BPS spectrum of theories realized by a 5-brane web with an O7-plane. In this paper, we propose such a method.

An insight for our proposal comes from rewriting the aforementioned SO($2N$) partition functions obtained in \cite{Hayashi:2020hhb, Nawata:2021dlk}, which are expressed as a Young diagram sum of factors including the ``Nekrasov factors'' and ``M-factors''. The Nekrasov factors are the typical terms that also appear in the U($N$) Nekrasov partition function. In contrast, the M-factors are not yet fully understood and their interpretation is less straightforward. A key technical point of this paper is to rewrite the M-factor in terms of a combination of Nekrasov factors. Our new expression indicates that newly written Nekrasov factors look consistent with the idea of the freezing realizing an  O7$^+$-plane as an O7$^-$-plane and (frozen) eight D7 branes discussed in \cite{Hayashi:2023boy, Kim:2024vci, Kim:2023qwh}. The rewriting of the M-factor also plays an important role in formulating the resultant partition function as a Young diagram sum.

As demonstrated in \cite{Hayashi:2023boy, Kim:2024vci, Kim:2023qwh}, the freezing is an effective way of obtaining various physical observables for the theories involving an O7$^+$-plane, which include the Seiberg-Witten curves, partition functions, and superconformal indices. For the freezing, eight D7-branes are introduced to an O7$^-$-plane  to make the RR-charge consistent. It is also worthy of noting that the combination of O7$^-$-plane with four D7-branes makes the RR-charge neutral, thereby realizing a $\mathbb{Z}_2$ orbifold, which was used in the context of S-folding \cite{Kim:2021fxx}. Combining these, we propose a novel idea that an O7$^+$-plane can be regarded as a $\mathbb{Z}_2$-orbifold with four D7-branes. We implement this to formulate a topological vertex for the SO($2N$) gauge theory based on a 5-brane web with an O7$^+$-plane.

The organization of the paper is as follows. In section \ref{sec:SO-via-O5}, we use the ``O-vertex'' proposed in \cite{Hayashi:2020hhb} to compute the partition function for SO($2N$) gauge theory, which is expressed in terms of ``M-factor'' as given in \cite{Nawata:2021dlk}. We introduce an identity associated with the M-factor, enabling us to rewrite the M-factor in terms of the Nekrasov factors. We then discuss an interpretation for this partition function from the point of view of the freezing \cite{Hayashi:2023boy,Kim:2024vci}. Based on the interpretation of the partition function as the freezing, we propose a topological vertex formalism with an O7$^+$-plane and show that our proposal indeed reproduces the SO($2N$) partition function in section \ref{sec:SO-via-O7}. We then conclude with a summary of our result and some generalizations in section \ref{sec:conclusion}. In appendices, we introduce our convention/notation, and also list relevant identities that the Nekrasov factor satisfies. Two different proofs for the M-factor identity are presented in the last appendix.

\section{SO partition function via 5-brane webs with an O5-plane}\label{sec:SO-via-O5}

In this section, we review how to compute the partition function of SO($2N$) gauge theory based on the O-vertex. We also discuss a key identity in performing the partition function computation and then present a new expression for the partition function of SO($2N$) gauge theory, which provides a new perspective on the topological vertex with an O7-plane. 

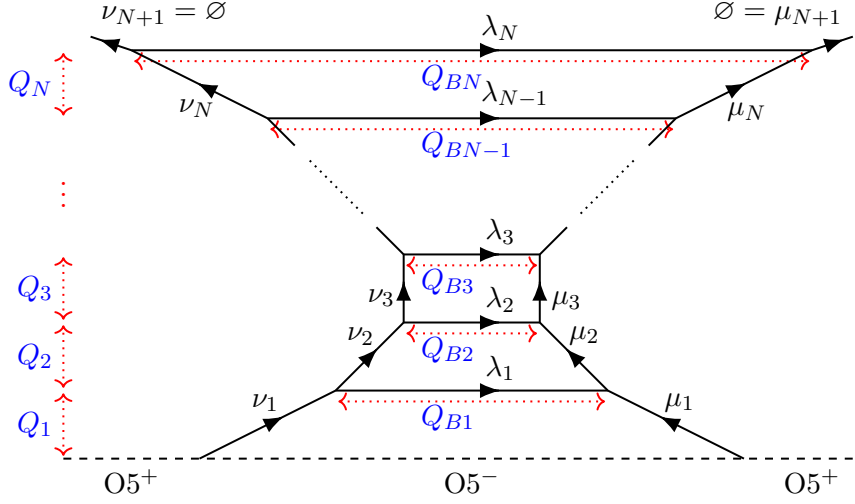
\begin{figure}[t]
    \centering
\begin{tikzpicture}[scale=1.8]
\draw [dashed] (-1,0)--(4.8,0);
\node at (2,0) [below] {O5${}^-$};
\node at (4.5,0) [below] {O5${}^+$};
\node at (-0.5,0) [below] {O5${}^+$};
\draw (1,0.5)  -- (3,0.5) ;
\draw (1.5,1)  -- (2.5,1);
\draw (1.5,1.5)-- (2.5,1.5);
\draw (0.5,2.5)-- (3.5,2.5);
\draw (-0.5,3) -- (4.5,3);
\draw [->, >={Latex[scale=1.2]}] (2.2,0.5)--(2.21,0.5)  
node [anchor=south]{$\lambda_1$};
\draw [->, >={Latex[scale=1.2]}] (2.2,1)  --(2.21,1) 
node [anchor=south]{$\lambda_2$};
\draw [->, >={Latex[scale=1.2]}] (2.2,1.5)--(2.21,1.5) 
node [anchor=south]{$\lambda_3$};
\draw [->, >={Latex[scale=1.2]}] (2.2,2.5)--(2.21,2.5) 
node [anchor=south]{$\quad \lambda_{N-1}$};
\draw [->, >={Latex[scale=1.2]}] (2.2,3)  --(2.21,3) 
node [anchor=south]{$\lambda_{N}$};
\draw[red,<->,dotted] (1.02,0.42)--(2.98,0.42);
\node at (1.55,0.3) [blue, right]{$Q_{B1}$};
\draw[red,<->,dotted] (1.52,0.92)--(2.48,0.92);
\node at (1.55,0.8) [blue, right]{$Q_{B2}$};
\draw[red,<->,dotted] (1.52,1.42)--(2.48,1.42);
\node at (1.55,1.3) [blue, right]{$Q_{B3}$};
\draw[red,<->,dotted] (0.52,2.42)--(3.48,2.42);
\node at (1.55,2.3) [blue, right]{$Q_{BN-1}$};
\draw[red, <->,dotted] (-0.48,2.92)--(4.48,2.92);
\node at (1.55,2.8) [blue, right]{$Q_{BN}$};
\draw (0,0)--(1,0.5)--(1.5,1)--(1.5,1.5)--(1.3,1.7);
\draw[dotted] (1.2,1.8) -- (0.8,2.2);
\draw (0.7,2.3) -- (0.5,2.5)--(-0.5,3)--(-0.8,3.1);
\draw [->, >={Latex[scale=1.2]}] (0.6,0.3) -- (0.62,0.31) 
node [anchor=south east][inner sep=1.5pt]{$\nu_1$};
\draw [->, >={Latex[scale=1.2]}] (1.3,0.8) -- (1.31,0.81) 
node [anchor=south east][inner sep=1pt]{$\nu_2$};
\draw [->, >={Latex[scale=1.2]}] (1.5,1.3) -- (1.5,1.31) 
node [anchor=north east][inner sep=3pt]{$\nu_3$};
\draw [->, >={Latex[scale=1.2]}] (0,2.75) -- (-0.02,2.76) 
node [anchor=north][inner sep=6pt]{$\nu_N$};
\draw [->, >={Latex[scale=1.2]}] (-0.71,3.07) -- (-0.74,3.08) ;
\node at (-0.8,3.1) [above right] {$\nu_{N+1} = \varnothing$} ;
\draw (4,0)--(3,0.5)--(2.5,1)--(2.5,1.5)--(2.7,1.7);
\draw[dotted] (2.8,1.8) -- (3.2,2.2);
\draw (3.3,2.3) -- (3.5,2.5)--(4.5,3)--(4.8,3.1);
\draw [->, >={Latex[scale=1.2]}] (3.4,0.3) -- (3.38,0.31) 
node [anchor=south west][inner sep=1.5pt]{$\mu_1$};
\draw [->, >={Latex[scale=1.2]}] (2.7,0.8) -- (2.69,0.81) 
node [anchor=south west][inner sep=1pt]{$\mu_2$};
\draw [->, >={Latex[scale=1.2]}] (2.5,1.3) -- (2.5,1.31) 
node [anchor=north west][inner sep=4pt]{$\mu_3$};
\draw [->, >={Latex[scale=1.2]}] (4,2.75) -- (4.02,2.76) 
node [anchor=north][inner sep=7pt]{$\mu_N$};
\draw [->, >={Latex[scale=1.2]}] (4.71,3.07) -- (4.74,3.08) ;
\node at (4.8,3.1) [above left] {$\varnothing = \mu_{N+1}$} ;
\draw[red,<->,dotted] (-1,0.02)--(-1,0.48);
\node at (-1,0.25) [blue, left] {$Q_1$};
\draw[red,<->,dotted] (-1,0.52)--(-1,0.98);
\node at (-1,0.75) [blue, left] {$Q_2$};
\draw[red,<->,dotted] (-1,1.02)--(-1,1.48);
\node at (-1,1.25) [blue, left] {$Q_3$};
\draw[red,<->,dotted] (-1,2.52)--(-1,2.98);
\node at (-1,2.75) [blue, left] {$Q_{N}$};
\node at (-1,2) [red] {$\vdots$};
\end{tikzpicture}
\caption{A 5-brane web for pure SO($2N$) gauge theory. An O5-plane is denoted by a dashed line at the bottom. We denote the K\"ahler parameters associated with each edge by $Q_i$ or $Q_{Bi}$ (painted in blue) and also denote Young diagrams by $\mu_i$, $\nu_i$, and $\lambda_i$.}
\label{fig:webForSO(2N)}
\end{figure}

\subsection{O-vertex}\label{sec:O-vertex}
Let us consider a 5-brane web diagram with an O5-plane which describes pure SO($2N$) gauge theory, as shown in Figure \ref{fig:webForSO(2N)}. By pure SO($2N$), we mean that the SO($2N$) gauge theory has no hypermultiplets. This means that there are no external D5-branes (or D7-branes) in the corresponding 5-brane web diagrams, as given in Figure \ref{fig:webForSO(2N)}. To account for an SO($2N$) gauge theory, an O5-plane is introduced in a 5-brane web, which is expanded along the same world volume direction as (color) D5-branes. As in Figure \ref{fig:webForSO(2N)}, an O5-plane is placed at the bottom and denoted by a dotted line. When an NS5-brane ends on an O5-plane, the RR charge of the O5-plane is altered and therefore the NS5-brane is titled such that the total charge is preserved at the intersection point, that is the vertex point where an NS5-brane ends on an O5-plane. For computation use, we introduce the parameters of the theory as follows: The K\"ahler parameters associated with the Coulomb branch are denoted by $Q_i$ where $i=1,\cdots, N$, while the K\"ahler parameter for the instanton fugacity is denoted by $Q_I$. 

\begin{figure}[H]
\centering
\begin{tikzpicture}
\draw [dashed] (-2,0)--(2,0);
\draw (0,0)--(1,0.5);
\node at (1,0.5) [above] {$\nu$};
\node at (1,0) [below] {O5${}^-$};
\node at (-1,0) [below] {O5${}^+$};
\node at (0,0.35) {$V_\nu$};
\end{tikzpicture}
\caption{O-vertex $V_{\nu}$ at the intersection between an O5-plane and a $(2,1)$ 5-brane associated with an edge of Young diagram $\nu$.}
\label{fig:o-vert}
\end{figure}
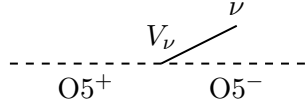

For a given 5-brane web diagram, one can compute the instanton partition function as an expansion of the instanton fugacity $Q_I$ \cite{Kim:2017jqn, Hayashi:2018bkd, Hayashi:2020hhb,Kim:2021cua}. It has been proposed \cite{Hayashi:2020hhb} that at least in the unrefined limit, where the Omega deformation parameters are set to $\epsilon_1=-\epsilon_2:= \hbar$, the instanton partition function of SO($2N$) gauge theory can be computed from the topological vertex formalism by introducing a new vertex for the intersection point between the orientifold plane and a $(2,1)$ 5-brane. More precisely, one assigns a newly introduced topological vertex $V_\nu$, called the O-vertex, as depicted in Figure~\ref{fig:o-vert}. 

The new vertex $V_\nu$ is defined as the expectation value, 
\begin{equation}
 V_\nu = (-1)^{|\nu|} \bra{0}\mathbb{O}(q)\ket{\nu}\ , \label{eq:O-vertexWithNuandP}
\end{equation}
where $q= e^{-\hbar}$ is the fugacity for the Omega deformation parameter and  the basis $\ket{\nu}$ is a ket of Frobenius basis labeled by the Young diagram $\nu$ satisfying the completeness relation of the basis,\footnote{
We note that our definitions of $V_\nu$ and $\mathbb{O}(q)$ are slightly different from those used in \cite{Nawata:2021dlk}. They are, in fact,  equivalent. This can be seen from properly rescaling $J_n$ by the associated K\"ahler parameter and appearing in \cite{Nawata:2021dlk}, $J_n \to A_1^{-n} J_n$. 
} 
\begin{equation}
    \sum_\nu\ket{\nu}\!\bra{\nu}=\mathbbm{1}\ .\label{eq:comp}
\end{equation}
The vertex operator $\mathbb{O}(q)$ in \eqref{eq:O-vertexWithNuandP} is defined as
\begin{equation}\label{Overtex}
\mathbb{O}(q) := \exp \lt(\sum_{n=1}^{\infty}-\frac{1+q^{n}}{2 n(1-q^{n})} J_{2 n}+\frac{1}{2 n} J_{n} J_{n}\rt) \ , 
\end{equation}
where $J_n$'s are free boson oscillators satisfying 
\begin{equation}\label{eq:Commutation-J}
    \lt[J_n,J_m\rt]=n\,\delta_{n+m,0}~.
\end{equation}
This vertex operator $\mathbb{O}(q)$ is determined so that the perturbative part of the partition function computed from the topological vertex formalism reproduces the expected form, as we will show it explicitly. 

The topological string partition function for pure SO($2N$) gauge theory is computed by gluing the left strip diagram $Z_{\text{left}}$ and the right strip diagram $Z_{\text{right}}$ of the web shown in Figure \ref{fig:webForSO(2N)}, which structurally leads to the following form:\footnote{If there are parallel edges in a given web diagram, there exist contributions that are independent of the Coulomb branch parameters. They are stringy contributions that should be removed to obtain the partition function for supersymmetric gauge theory. For pure SO($2N$) gauge theory, it happens when $N=2$. Therefore, it is necessary to remove this extra factor from this expression given in \eqref{eq:glueLR}. 
In the following, we assume $N \ge 3$.}
\begin{align}\label{eq:glueLR}
    Z_{{\rm SO}(2N)}^{\textrm{top}}
&=\sum_{\vec{\lambda}} Z_{\text{left}} Z_{\text{right}}
\prod_{s=1}^N (-Q_{B s})^{|\lambda_{s}|} f_{\lambda_{s}}^{5-2s} \ , 
\end{align}
where $Q_{B s}$ is the K\"ahler parameters corresponding to the horizontal edges or color D5-branes, which are given in terms of the instanton factor $Q_I$ as
\begin{align}\label{eq:ParamQB}
Q_{B s} & = (-1)^{N} Q_I\, A_s^{2s-4} \prod_{r=s+1}^N A_r^2\quad \qquad (s=1,\cdots, N)\ ,
\end{align}
with
\begin{equation}
    A_s:=\prod_{t=1}^s Q_t\ .
\end{equation}
Here, the left strip and the right strip are represented in terms of the O-vertex, respectively as
\begin{align}
Z_{\text{left}} 
&= \sum_{\vec{\nu}} 
\lt(\prod_{r=1}^N(-Q_r)^{|\nu_{r}|}\prod_{s=2}^Nf_{\nu_{s}}^{-1}\rt)
V_{\nu_{1}}
\prod_{s=1}^N C_{\nu_{s}^T \nu_{s+1} \lambda_{s}}\ ,
\cr
Z_{\text{right}} 
&= \sum_{\vec{\mu}} 
\lt(\prod_{r=1}^N(-Q_r)^{|\mu_{r}|}\prod_{s=2}^Nf_{\mu_{s}}\rt)
W_{\mu_{1}} 
  \prod_{s=1}^N C_{\mu_{s+1} \mu_{s}^T \lambda_{s}^T}  \ ,
\end{align}
where the superscript ${}^T$ denotes the transpose of Young diagram,  we set $\nu_{s+1}=\mu_{s+1}=\varnothing$, and $W_{\mu} = q^{\frac{\kappa(\mu)}{2}} V_{\mu^T}$. Here, the framing factor $f_\lambda$ takes the form  
\begin{equation}
    f_\lambda:=(-1)^{|\lambda|}q^{\frac{\kappa(\lambda)}{2}},\qquad \kappa(\lambda):=2\sum_{(i,j)\in\lambda}(j-i)\ , \label{framing}
\end{equation}
and the vertex factor $C_{\nu\mu\lambda}$ can be expressed with (skew) Schur functions as 
\begin{align}
    C_{\nu\mu\lambda}
& =q^{\frac{1}{2}\big(\kappa(\mu)+\kappa(\lambda)\big)}
s_{\lambda}(q^{-\rho})\sum_{\sigma}s_{\nu^T/\sigma}(q^{-\rho-\lambda})s_{\mu/\sigma}(q^{-\rho-\lambda^T})\ .
\label{topvertex}
\end{align}

One can evaluate the left/right strip contributions by applying the Cauchy identity,
\begin{align}
& \sum_\lambda Q^{|\lambda|}\, s_{\lambda/\sigma}(q^{-\rho-\mu})\,s_{\lambda/\tau}(q^{-\mu-\nu})
\cr
&~~= R^{-1}_{\mu \nu}(Q;q) \!\sum_\eta Q^{|\sigma|+|\tau|-|\eta|}\,s_{\tau/\eta}(q^{-\rho-\mu})\,s_{\sigma/\eta}(q^{-\rho-\nu})\ ,\label{Schur-id-spec}
\end{align}
where 
\begin{align}\label{eq:def-R-N}
R_{\mu \nu} (Q;q) 
&= \prod_{i=1}^{\infty} \prod_{j=1}^{\infty}
\lt( 1 - Q q^{i+j-1-(\mu)_i-(\nu)_j} \rt)= \mathrm{PE} \lt( - \frac{q}{(1-q)^2} Q \rt) {N}_{\mu^T \nu} (Q;q)\ , 
\end{align}
with PE being the Plethystic exponential and $N_{\mu^T\nu}(Q; q)$ being the Nekrasov factor defined by
\begin{align}
     N_{\mu^T \nu} (Q;q) :=\prod_{(i,j)\in \mu}\lt(1-Qq^{1-i-j+(\mu^T)_j+(\nu^T)_i}\rt)\prod_{(i,j)\in \nu}\lt(1-Qq^{i+j-(\mu)_j-(\nu)_i-1}\rt) \ .
     \label{eq:N-factor}
\end{align}
(See Appendix \ref{App:B} for its properties.) One finds, for example, the contributions from the left strip diagram are given by
\begin{align}
Z_{\text{left}}
= & 
\sum_{\nu_{1},\eta_2, \cdots, \eta_N} \! (-Q_1)^{|\nu_{1}|}\! \prod_{r=2}^N\!Q_r^{|\eta_{r}|}\! \times \! V_{\nu_{1}}\,
\prod_{s=1}^N q^{\frac{\kappa(\lambda_s)}{2}} s_{\lambda_s}(q^{-\rho}) \times \prod_{s<t} R^{-1}_{\lambda_{s}^T \lambda_{t}}(A_t/A_s;q)
\cr
& \quad
\times 
s_{\nu_1/\eta_2}(q^{-\rho-\lambda_{1}})s_{\eta_2/\eta_3}(q^{-\rho-\lambda_{2}})\cdots 
s_{\eta_{N-1}/\eta_{N}}(q^{-\rho-\lambda_{N-1}})
s_{\eta_{N}}(q^{-\rho-\lambda_{N}})
\end{align}
In this expression, we can absorb the K\"ahler parameters into the skew Schur functions by using the identity
\begin{align}
Q^{|\lambda|-|\mu|}  s_{\lambda/\mu}(\vec{x}) = s_{\lambda/\mu}(Q \vec{x})
\end{align}
as
\begin{align}
Z_{\text{left}}
= &   
\prod_{s=1}^N s_{\lambda_s^T}(q^{-\rho}) 
\times \prod_{s<t} R^{-1}_{\lambda_{s}^T \lambda_{t}}(A_t/A_s;q) 
\times 
\sum_{\vec{\eta}} (-1)^{|\eta_1|} V_{\eta_{1}} \prod_{s=1}^{N} \,s_{\eta_s/\eta_{s+1}}(q^{-\rho-\lambda_{s}} A_s) \ ,
\label{M-strip-left}
\end{align}
where we denote $\vec{\eta} = (\eta_1 ,\cdots, \eta_{N})$ by relabeling $\nu_1 \to \eta_1$ and set $\eta_{N+1}=\varnothing$. We have also used the identity $q^{\frac{\kappa(\lambda_s)}{2}} s_{\lambda_s}(q^{-\rho}) = s_{\lambda_s^T}(q^{-\rho})$.
We note that the slew Schur functions can be rewritten as an operator expression, 
\begin{align}
 s_{\lambda/\mu}(\vec{x})=\bra{\mu}\Gamma_+(\vec{x})\ket{\lambda}=\bra{\lambda}\Gamma_-(\vec{x})\ket{\mu}  \ ,  
\end{align} 
with
\begin{equation}\label{eq:Gamma-pm}
    \Gamma_\pm (\vec{x}):=\exp\lt(\sum_{n=1}^\infty \frac{1}{n}\sum_i x_i^n J_{\pm n}\rt)~. 
\end{equation}
It follows from the completeness relation of the basis \eqref{eq:comp} that the left strip part can be given in a compact form as 
\begin{align}
Z_{\text{left}}
&=  \prod_{s=1}^N s_{\lambda_s^T}(q^{-\rho}) \cdot
\prod_{s<t}R^{-1}_{\lambda_{s}^T \lambda_{t}}(A_t A^{-1}_s;q) \cdot \bra{0}\mathbb{O}(q)\prod_{t=1}^N\Gamma_-(A_tq^{-\rho-\lambda_{t}})\ket{0}\ .
\label{M-strip}
\end{align}
A similar computation applies to the right strip part $Z_{\rm right}$ and the two strip contributions are related as follows: 
\begin{align}
\frac{Z_{\text{right}}}{\displaystyle\prod_{s=1}^N  s_{\lambda_s}(q^{-\rho}) } 
= \frac{Z_{\text{left}}}{\displaystyle\prod_{s=1}^N s_{\lambda_s^T}(q^{-\rho}) }\ .
\label{eq:ZL=ZR}
\end{align}

We now equate the topological string partition function $Z_{{\rm SO}(2N)}^{\textrm{top}}$ obtained from topological vertex \eqref{eq:glueLR} to the supersymmetric gauge theory partition function which can be given in a product form in terms of the perturbative and instanton part, {\it i.e.}, 
\begin{equation}
Z_{{\rm SO}(2N)}^{\textrm{top}} = {Z_{{\rm SO}(2N)}^{\textrm{pert}}} Z_{{\rm SO}(2N)}^{\textrm{inst}} \ .\label{instanton-part}
\end{equation}
Here, the perturbative part ${Z_{{\rm SO}(2N)}^{\textrm{pert}}}$ is obtained by taking the limit $Q_I \to 0$ of the topological string partition function $Z_{{\rm SO}(2N)}^{\textrm{top}}$ given in \eqref{eq:glueLR}.  It follows from \eqref{M-strip} and \eqref{eq:ZL=ZR} that the perturbative part takes the form
\begin{align}\label{eq:top-Qto0}
\lim_{Q_I \to 0} Z_{{\rm SO}(2N)}^{\textrm{top}}
& = Z_{\text{left}} Z_{\text{right}} \Big|_{\vec{\lambda} = \vec{\varnothing}}
\cr
&
= \prod_{s<t}{\rm PE}\lt(\frac{2q}{(1-q)^2}\frac{A_t}{A_s}\rt)\!
\bigg(
\bra{0}\mathbb{O}(q)\prod_{r=1}^N\Gamma_-(A_r \,q^{-\rho})\ket{0}
\bigg)^2 .
\end{align}
Since the factor including the O-vertex in \eqref{eq:top-Qto0} yields \cite{Hayashi:2020hhb}
\begin{align}\label{eq:expectO-empty}
\bra{0}\mathbb{O}(q)\prod_{t=1}^N\Gamma_-(A_t\,q^{-\rho})\ket{0}
&= 
{\rm PE}\Bigg(\frac{q}{2(1-q)^2} \bigg( \Big(\sum_{t=1}^N A_t \Big)^2 - \sum_{t=1}^N A_t{}^2 \bigg) \Bigg)
\cr
&= 
{\rm PE}\lt(\frac{q}{(1-q)^2} \sum_{s<t} A_t A_s \rt), 
\end{align}
one easily finds that \eqref{eq:top-Qto0} indeed reproduce the expected form of the perturbative part
\begin{align}
    Z_{{\rm  SO}(2N)}^{\textrm{pert}}&=\textrm{PE}\Bigg(\frac{2q}{(1-q)^2}\sum_{\alpha\in\Delta_+}e^{-\alpha\cdot a }\Bigg)
    =\textrm{PE}\lt( \frac{2q}{(1-q)^2}\sum_{s<t}^N \lt( A_tA_s +A_tA_s^{-1}   \rt)\rt) ,
    \label{pert-part}
\end{align}
where $\Delta_+$ denotes the positive roots of the Lie algebra of SO($2N$). 

Let us compute the instanton part ${Z_{{\rm SO}(2N)}^{\textrm{inst}}}$, which should be obtained by dividing the topological string partition function by the perturbative part, as given in \eqref{instanton-part}, where $Z_{{\rm SO}(2N)}^{\textrm{top}}$ is obtained from \eqref{M-strip} and \eqref{eq:ZL=ZR}. A little calculation leads to
\begin{align}\label{eq:SO-via-M}
    Z_{{\rm SO}(2N)}^{\textrm{inst}}
=& \sum_{\vec{\lambda}} 
\prod_{s=1}^N (-Q_{B s})^{|\lambda_{s}|} f_{\lambda_{s}}^{5-2s}
s_{\lambda_s}(q^{-\rho}) s_{\lambda_s^T}(q^{-\rho})
\, \times\!\!
\prod_{1\le s< t\le N}  N^{-2}_{\lambda_{s}\lambda_{t}}(A_{t}A_s^{-1};q)
\cr
 & \quad  \times M^2_{\vec{\lambda}}(A_1,A_2,\cdots,A_N)\ , 
\end{align}
where we collectively introduce Young diagrams as $\vec{\lambda} = (\lambda_1, \lambda_2,\cdots, \lambda_N)$ and 
\begin{align}\label{eq:M-factor-op-text}
M_{\vec{\lambda}}(A_1,A_2,\cdots,A_N)
:= 
\frac{\bra{0}\mathbb{O}(q)\prod_{s=1}^N \Gamma_-(A_s q^{-\rho-\lambda_{s}})\ket{0}}{\bra{0}\mathbb{O}(q)\prod_{s=1}^N \Gamma_-(A_sq^{-\rho})\ket{0}}\ ,
\end{align}
which we refer to as the ``{\it M-factor}'' from here on. We note that the M-factor provides a new perspective on reformulating the O-vertex which we shall shortly discuss.

\subsection{M-factor identity and New expression for SO partition function}
We propose the square of this M-factor can be expressed in terms of the Nekrasov factors, which we refer to as the M-factor identity:
\begin{equation} 
 \boxed{
 \begin{aligned}
 M^2_{\vec{\lambda}}(A_1,\cdots,A_N)
  &=
 \prod_{s=1}^N \prod_{t=1}^N 
 N^{-1}_{\lambda_{s}^T \lambda_t}(A_s A_t;q)
 \times \prod_{r=1}^N \prod_{f=1}^4 \prod_{\ell=\pm 1}
 N_{\lambda_{r}^T \varnothing} (A_{r} \,{\sf M}_f^{\ell} ;q)\ , 
 \label{eq:M-factor}
\end{aligned}
}
\end{equation}
where 
\begin{align}\label{eq:defM1234}
{\sf M}_1 := 1, \quad
{\sf M}_2 := - 1, \quad
{\sf M}_3 := q^{\frac12}, \quad {\rm and}\quad  
{\sf M}_4 := - q^{\frac12}\ .
\end{align}
The proof of the M-factor identity is presented in appendix \ref{App:C}. 

Implementing the M-factor identity \eqref{eq:M-factor}, we re-express the SO($2N$) Nekrasov instanton partition function \eqref{eq:SO-via-M} in terms of the Nekrasov factors: 
\begin{align}\label{eq:SO-inst-1}
    Z_{{\rm SO}(2N)}^{\textrm{inst}}
=& \sum_{\vec{\lambda}} \prod_{s=1}^N (-Q_{B s})^{|\lambda_{s}|} f^{5-2s}_{\lambda_{s}} 
s_{\lambda_s}(q^{-\rho}) s_{\lambda_s^T}(q^{-\rho})
\times  
\prod_{1\le s< t\le N} \!\!\! N^{-2}_{\lambda_{s}\lambda_{t}}(A_{t}A_s^{-1};q)
\cr
 & \qquad \times 
\prod_{s=1}^N \prod_{t=1}^N 
N^{-1}_{\lambda_{s}^T \lambda_t}(A_s A_t;q)
\times \prod_{r=1}^N \prod_{f=1}^4 \prod_{\ell=\pm 1}\!\!
N_{\lambda_{r}^T \varnothing} (A_{r}\, {\sf M}_f^{\ell} ;q)\ .
\end{align}
We can use the identities listed in Appendix \ref{App:B} to further simplify the partition function. For instance, the following identity 
\begin{align}
N^{-1}_{\lambda_s\lambda_t}(Q;q)
&=
(-Q^{-1})^{|\lambda_s|+|\lambda_t|}q^{\frac{1}{2}\kappa(\lambda_t)-\frac{1}{2}\kappa(\lambda_s)} N^{-1}_{\lambda_t\lambda_s} \lt( Q^{-1};q \rt)\ ,
\end{align}
yields 
\begin{align}\label{eq:Nst-inv}
&\prod_{1 \le s < t \le N} N^{-2}_{\lambda_s \lambda_t}(A_t A_s^{-1};q)
\cr
&\quad = \prod_{1 \le s < t \le N} 
(-1)^{|\lambda_s|+|\lambda_t|}
(A_s A_t^{-1})^{|\lambda_s|+|\lambda_t|}
q^{\frac{1}{2}(\kappa(\lambda_t)-\kappa(\lambda_s))}
N_{\lambda_t \lambda_s}^{-1} (A_t^{-1} A_s;q)
\cr
& \qquad \times \prod_{1 \le s < t \le N} N^{-1}_{\lambda_s \lambda_t}(A_t A_s^{-1};q)
\cr
&\quad= 
\prod_{s=1}^N
(-1)^{(N-1)|\lambda_s|}
\bigg( A_s^{N-2s} \prod_{r=1}^N A_r \prod_{r=s+1}^N A_r^{-2}
\bigg)^{|\lambda_s|}
q^{-\frac{1}{2}(N-2s+1) \kappa(\lambda_s)}
\cr
& \qquad \times
\prod_{s \neq t} N^{-1}_{\lambda_s \lambda_t}(A_t A_s^{-1};q)\ , 
\end{align}
where in the second equality, we have used 
\begin{align}
&\prod_{1\le s<t \le N}
q^{\frac{1}{2} \left( \kappa(\lambda_t) - \kappa(\lambda_s) \right)}
= 
\prod_{s=1}^N q^{-\frac{1}{2}(N-2s+1) \kappa(\lambda_s)}\ , 
\cr
&
\prod_{1\le s<t \le N} (-1)^{|\lambda_s|+|\lambda_t|}
= \prod_{s=1}^N (-1)^{(N-1)|\lambda_s|}\ , 
\cr
& 
\prod_{1\le s<t \le N} (A_s A_t^{-1})^{|\lambda_s|+|\lambda_t|}
= \prod_{s=1}^N \bigg( A_s^{N-2s} \prod_{r=1}^N A_r \prod_{r=s+1}^N A_r{}^{-2}
\bigg)^{|\lambda_s|} \ .
\end{align}
Also by substituting \eqref{eq:ParamQB}, we find that \eqref{eq:SO-inst-1} is rewritten as
\begin{align}\label{eq:5dSO-inst-middle}
    Z_{{\rm SO}(2N)}^{\textrm{inst}}
=& \sum_{\vec{\lambda}} \prod_{s=1}^N \left( Q_I A_s^{N-4} \prod_{r=1}^N A_r \right)^{|\lambda_{s}|} 
q^{\frac{(4-N)\kappa(\lambda_s)}{2}}
\\
&  
\times 
\prod_{s, t=1}^N 
N_{\lambda_{s} \lambda_{t}}^{-1}(A_{t}A_s^{-1};q)\,
N_{\lambda_{s}^T \lambda_t}^{-1}(A_s A_t;q)
\times \prod_{r=1}^N \prod_{f=1}^4 \prod_{\ell=\pm 1}
N_{\lambda_{r}^T \varnothing} (A_r {\sf M}_f{}^{\ell} ;q)\ .\nonumber
\end{align}
This expression becomes even simpler by introducing a new Nekrasov factor defined as 
\begin{align}\label{eq:def-nekt}
\tilde{n}_{\lambda\nu}(a;\hbar) 
&:= 
A^{-\frac12 \big(|\lambda|+|\nu|\big)} q^{-\frac14 \big(\kappa(\lambda) - \kappa(\nu)\big)}
N_{\lambda\nu}(A;q) 
\cr
&= \prod_{(i,j)\in \lambda}
2\, \sinh\frac12 \Big(a + \hbar \big(1-i-j+(\lambda)_i+(\nu^T)_j \big)\Big)
\cr
& \quad \times
\prod_{(i,j)\in \nu}
2 \,\sinh \frac12 \Big(a + \hbar \big(i+j-1-(\lambda^T)_j-(\nu)_i \big)\Big)\ ,
\end{align}
where $A=e^{-a}$ and $q=e^{-\hbar}$, and they satisfy
\begin{align}\label{eq:newNekr-rel}
\tilde{n}_{\lambda\nu}(a;\hbar)
= \tilde{n}_{\nu^T \lambda^T}(a;\hbar)
= (-1)^{|\lambda|+|\nu|} \tilde{n}_{\nu\lambda}(-a;\hbar)\ .
\end{align} 
See Appendix \ref{App:B} for more details. It follows that each factor in \eqref{eq:5dSO-inst-middle} can be rewritten in terms of the new Nekrasov factor as
\begin{align}
\prod_{s=1}^N \prod_{t=1}^N 
N^{-1}_{\lambda_{s} \lambda_{t}}(A_{t}A_s^{-1};q)
&= \prod_{s=1}^N \prod_{t=1}^N 
\tilde{n}^{-1}_{\lambda_{s} \lambda_{t}}(a_t - a_s ;\hbar ) \ ,
\cr
\prod_{s=1}^N \prod_{t=1}^N 
N^{-1}_{\lambda_{s}^T \lambda_{t}}(A_t A_s;q)
&= 
\prod_{s=1}^N  
\bigg( A_s^{-N} \prod_{r=1}^N A_r^{-1} \bigg)^{|\lambda_s|}
q^{\frac{1}{2}N \kappa(\lambda_s)} 
\times
\prod_{s=1}^N \prod_{t=1}^N 
\tilde{n}^{-1}_{\lambda_{s}^T \lambda_{t}} (a_t + a_s;\hbar) \ ,
\cr
\prod_{f=1}^4 \prod_{\ell=\pm 1}
N_{\lambda_{s}^T \varnothing} (A_s {\sf M}_f^{n} ;q)
&=  
A_s^{4|\lambda_s|} q^{-2 \kappa(\lambda_s)}
\prod_{f=1}^4 \prod_{\ell=\pm 1}
\tilde{n}_{\lambda_{s}^T \varnothing} (a_s + \ell {\sf m}_f ;\hbar) \ ,
\end{align}
which leads to a useful expression for the Nekrasov instanton partition function for 5d $\mathcal{N}=1$ pure SO($2N$) gauge theory:
\begin{align}\label{eq:5D-SO-final}
    Z_{{\rm SO}(2N)}^{\textrm{inst}}
=& \sum_{\vec{\lambda}} Q_I^{|\vec{\lambda}|} 
\frac{\displaystyle
\prod_{s=1}^N \prod_{f=1}^4 \prod_{\ell =\pm 1} 
\tilde{n}_{\lambda_{s}^T \varnothing} (a_s + \ell {\sf m}_f;\hbar)
}{\displaystyle
\prod_{s=1}^N \prod_{t=1}^N 
\tilde{n}_{\lambda_{s} \lambda_{t}}(a_t-a_s;\hbar) \cdot
\tilde{n}_{\lambda_{s}^T \lambda_t}(a_t+a_s;\hbar)}\ ,
\end{align}
where
\begin{align}\label{eq:4d-mass}
&{\sf m}_1 := 0, \quad
{\sf m}_2 := \pi i, \quad
{\sf m}_3 := \frac12 \hbar, \quad
{\sf m}_4 := \frac12 \hbar + \pi i\ .
\end{align}
This is our new expression for the Nekrasov partition function, which is given in terms of the Young diagram sums of the combination of the new Nekrasov factors.

\begin{figure}[htbp]
    \centering
\begin{tikzpicture}[scale=1.1]
\draw [dashed] (-1,0)--(5,0);
\draw (0.5,0)--(1.5,0.5);
\draw (1.5,0.5)--(2.75,0.5);
\draw (1.5,0.5)--(2,1);
\draw (2,1)--(2.75,1);
\draw (2,1)--(2,1.5);
\draw (2,1.5)--(3.25,1.5);
\draw (2,1.5)--(1.5,2);
\draw (2.75,1)--(3.25,1.5);
\draw (3.25,1.5)--(4.25,2);
\draw (3,0.25)--(3.5,0);
\draw (3,0.25)--(2.25,0.25);
\draw (3,0.25)--(2.75,0.5);
\draw (2.75,0.5)--(2.75,1);
\draw[red, <->,dotted] (3.55,0)--(3.55,0.3);
\node at (3.55,0.15) [right] {$T \to 1$};
\node at (2.25,0.25) [left] {$\varnothing$};
\end{tikzpicture}
\caption{An example for the effective brane web of the pure SO(7) gauge theory. One can first compute the partition function for this diagram with the K\"ahler parameter $T$ keeping it finite and then take the limit $T\to 1$ at the end.}
\label{fig:SO7-brane}
\end{figure}
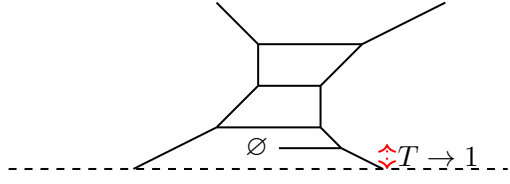

\paragraph{Remark:} 
Based on the discussion above, it is also straightforward to obtain the Nekrasov instanton partition function for SO($2N+1$) gauge theories.
One can apply the O-vertex formalism to the effective brane web shown in Figure \ref{fig:SO7-brane} \cite{Hayashi:2020hhb}. Using an identity of the M-factor \cite{Nawata:2021dlk}
\begin{equation}
\lim_{T \to 1} M_{\varnothing,\lambda_{1},\cdots,\lambda_{N}}(T,A_1,\cdots,A_{N})
= 
M_{\lambda_{1},\cdots,\lambda_{N}}(A_1,\cdots,A_{N})\prod_{s=1}^N N^{-1}_{\varnothing \lambda_{s}}(A_s ; q) \ ,
\end{equation}
one finds that the instanton partition function of pure SO($2N+1$) theory can be written as 
\begin{align}\label{eq:SOod-via-M}
    Z_{{\rm SO}(2N+1)}^{\textrm{inst}}
=& \sum_{\vec{\lambda}} 
\prod_{s=1}^N (-Q'_{B s})^{|\lambda_{s}|} f_{\lambda_{s}}^{4-2s}
\,s_{\lambda_s}(q^{-\rho})\, s_{\lambda_s^T}(q^{-\rho})
\,N^{-2}_{\varnothing \lambda_{s}}(A_s ; q)
\cr
 & \qquad 
\times
\prod_{1\le s< t\le N}  N^{-2}_{\lambda_{s}\lambda_{t}}(A_{t}A_s^{-1};q)
\times M^2_{\vec{\lambda}}(A_1,A_2,\cdots,A_N)\ ,
\end{align}
where the K\"ahler parameters corresponding to the horizontal edges are slightly modified from \eqref{eq:ParamQB} as 
\begin{align}\label{eq:QBIt}
Q'_{B_s} & = (-1)^N Q_I A_s^{2s-3} \prod_{r=s+1}^N A_r^2
\quad\qquad (s=1,\cdots, N)\ .
\end{align}

Rewriting \eqref{eq:SOod-via-M} using \eqref{eq:def-nekt} and \eqref{eq:QBIt} leads to 
\begin{align}\label{eq:5D-SOodd-final}
    Z_{{\rm SO}(2N+1)}^{\textrm{inst}}
=& \sum_{\vec{\lambda}} 
Q_I^{|\vec{\lambda}|} 
\frac{\displaystyle
\prod_{s=1}^N \prod_{f=2}^4 \prod_{\ell =\pm 1} 
\tilde{n}_{\lambda_{s}^T \varnothing} (a_s + \ell {\sf m}_f;\hbar)
}{\displaystyle
\prod_{s=1}^N \prod_{t=1}^N 
\tilde{n}_{\lambda_{s} \lambda_{t}}(a_t-a_s;\hbar) \cdot
\tilde{n}_{\lambda_{s}^T \lambda_t}(a_t+a_s;\hbar)}
\ .
\end{align}
Compared to the case for SO($2N$), the factor including ${\sf m}_1$ is missing in this case.

\paragraph{Remark:}The 4d limit is obtained from \eqref{eq:5D-SO-final} by replacing $a_s \to \beta a_s$, $\hbar \to \beta \hbar$ and by taking the limit $\beta \to 0$ while $a_s$ and $\hbar$ being finite. Note that, when $\beta \to 0$ while keeping $a_s$ finite,
\begin{align}
\tilde{n}_{\lambda \mu} (\beta a; \beta \hbar)
\sim \beta^{|\lambda|+|\mu|} \tilde{n}^{4d}_{\lambda \mu} (a;\hbar),
\qquad
\tilde{n}_{\lambda \mu} (\beta a \pm \pi i;\beta \hbar)
\sim 1,
\end{align}
where we defined the 4d Nekrasov factor as 
\begin{align}
\tilde{n}^{4d}_{\lambda\nu}(a;\hbar) 
:= &\prod_{(i,j)\in \lambda}
\!\!\lt(a + \hbar (1-i-j+(\lambda)_i+(\nu^T)_j )\rt)
\cdot\!\!\!
\prod_{(i,j)\in \nu}\!\!
\lt(a + \hbar (i+j-1-(\lambda^T)_i-(\nu)_j )\rt).
\end{align}
Introducing the 4d instanton factor $Q^{4d}$ as
\begin{align}
Q_I = \beta^{2N-8} Q^{4d}\ ,
\end{align}
we find Nekrasov partition function for 4d pure SO($2N$) gauge theory is given by
\begin{align}\label{eq:4D-SO-final}
    Z_{\text{4d SO}(2N)}^{\textrm{inst}}
=& \sum_{\vec{\lambda}} (Q^{4d})^{|\vec{\lambda}|} 
\frac{\displaystyle
\prod_{s=1}^N \prod_{f=1,3} \prod_{\ell=\pm 1}
\tilde{n}^{4d}_{\lambda_{s}^T \varnothing} (a_s + \ell {\sf m}_{f};\hbar)
}{\displaystyle
\prod_{s=1}^N \prod_{t=1}^N 
\tilde{n}^{4d}_{\lambda_{s} \lambda_{t}}(a_t-a_s;\hbar) \cdot
\tilde{n}^{4d}_{\lambda_{s}^T \lambda_t}(a_t+a_s;\hbar)}\ ,
\end{align}
where ${\sf m}_{f}$ $(f=1,3)$ are identical to the one given in \eqref{eq:4d-mass}.
Analogously, for 4d pure SO($2N+1$) gauge theory, we have
\begin{align}\label{eq:4D-SOodd-final}
    Z_{\text{4d SO}(2N+1)}^{\textrm{inst}}
=& \sum_{\vec{\lambda}} (Q^{4d})^{|\vec{\lambda}|} 
\frac{\displaystyle
\prod_{s=1}^N \prod_{\ell=\pm 1}
\tilde{n}^{4d}_{\lambda_{s}^T \varnothing} (a_s + \frac12 \ell \hbar;\hbar)
}{\displaystyle
\prod_{s=1}^N \prod_{t=1}^N 
\tilde{n}^{4d}_{\lambda_{s} \lambda_{t}}(a_t-a_s;\hbar) \cdot
\tilde{n}^{4d}_{\lambda_{s}^T \lambda_t}(a_t+a_s;\hbar)}\ .
\end{align}

\subsection{Interpretation as a freezing}\label{sec:int-freeze}

In this subsection, we give some interpretation for the new expression for the Nekrasov partition function in \eqref{eq:5D-SO-final} with \eqref{eq:4d-mass} based on the 5-brane web with an O5-plane.

A generic U($N$) (or SU($N$)) instanton partition is organized as a product of the contributions from the vector multiplet and the matter multiplets. The terms in the denominator are responsible for the vector contribution, while those in the numerator are responsible for the matter multiplets, the form of which changes depending on the representation of the matter multiplet. With this in mind, we regard the Nekrasov factor of the form $\tilde{n}_{\lambda_{s} \lambda_{t}}(a;\hbar)$ as the contribution from a string with the length $a$. 

\begin{figure}[H]
    \centering
\begin{minipage}{7cm}
\centering
\begin{tikzpicture}
\draw [dashed] (-1,0)--(4.8,0);
\node at (2,0) [below] {O5${}^-$};
\node at (4.5,0) [below] {O5${}^+$};
\node at (-0.5,0) [below] {O5${}^+$};
\draw (1,0.5)  -- (3,0.5) ;
\draw (1.5,1)  -- (2.5,1);
\draw (1.5,1.5)-- (2.5,1.5);
\draw (0,0)--(1,0.5)--(1.5,1)--(1.5,1.5)--(1.3,1.7);
\draw (4,0)--(3,0.5)--(2.5,1)--(2.5,1.5)--(2.7,1.7);
\draw [dotted] (1,0.5)  -- (-0.3,0.5) node [left] {$a_s$} ;
\draw [dotted] (1.5,1.5)-- (-0.3,1.5) node [left] {$a_t$} ;
\draw[decorate, blue, decoration={snake,segment length=6pt, amplitude=1.5pt}] (2,0.5) -- (2,1.5);
\end{tikzpicture}
\end{minipage}
\begin{minipage}{7cm}
\centering
\begin{tikzpicture}
\draw [dashed] (-1,0)--(4.8,0);
\node at (2,0) [below] {O5${}^-$};
\node at (4.5,0) [below] {O5${}^+$};
\node at (-0.5,0) [below] {O5${}^+$};
\draw (1,0.5)  -- (3,0.5) ;
\draw (1.5,1)  -- (2.5,1);
\draw (1.5,1.5)-- (2.5,1.5);
\draw (0,0)--(1,0.5)--(1.5,1)--(1.5,1.5)--(1.3,1.7);
\draw (4,0)--(3,0.5)--(2.5,1)--(2.5,1.5)--(2.7,1.7);
\draw [dotted] (1,0.5)  -- (-0.3,0.5) node [left] {$a_s$} ;
\draw [dotted] (1.5,1.5)-- (-0.3,1.5) node [left] {$a_t$} ;
\draw[decorate, red, decoration={snake,segment length=6pt, amplitude=1.5pt}] (2,0.04) -- (1.8,1.5);
\draw[decorate, red, decoration={snake,segment length=6pt, amplitude=1.5pt}] (2.25,0.5)-- (2.135,0.04) ;
\draw[red] (2,0.05) arc (-140:-40:0.09);
\end{tikzpicture}
\end{minipage}
\caption{Two types of strings that connect the $s$-th color brane and the $t$-th color brane. {\sf LEFT}: the string (in {\color{blue} blue}) connecting them directly. {\sf RIGHT}: the string (in {\color{red}red}) connecting them through O5-plane. The strings are denoted by the wavy lines. }
\label{fig:O5-interpret}
\end{figure}
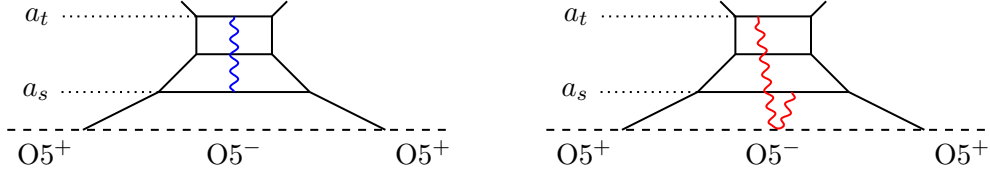

Consider the two factors in the denominator in \eqref{eq:5D-SO-final}. The first factor
\begin{align}\label{eq:firstfactorInDenominator}
\tilde{n}_{\lambda_{s} \lambda_{t}}(a_t-a_s;\hbar) 
\end{align}
is interpreted as the contribution from the string directly connecting the $s$-th color D5-brane at the height $a_s$ and the $t$-th color D5-brane at the height $a_t$ where both heights are measured from the position of the O5-plane, as depicted on the left-hand side of Figure \ref{fig:O5-interpret}. The second factor in the denominator
\begin{align}\label{eq:2ndfactorInDenominator}
\tilde{n}_{\lambda_{s}^T \lambda_t}(a_t+a_s;\hbar)    
\end{align}
is interpreted as the contribution from the string connecting the $s$-th color D5-brane and the $t$-th color D5-brane through the O5-plane, as depicted on the right-hand side of Figure~\ref{fig:O5-interpret}.

\noindent This interpretation for the factors in the denominator naturally suggests that, at least for $t \neq s$, they are interpreted as the W-bosons of the SO($2N$) gauge theory. The first factor \eqref{eq:firstfactorInDenominator} with $t=s$ would then correspond to the Cartan part of the SO($2N$). On the other hand, the second factor \eqref{eq:2ndfactorInDenominator} with $t=s$ does not seem to have a clear interpretation. We note however that this term corresponds to the term originated from the diagonal part given in \eqref{eq:M-diag-id}. It would be still good to find an intuitive interpretation in terms of fundamental strings connecting color D-branes.

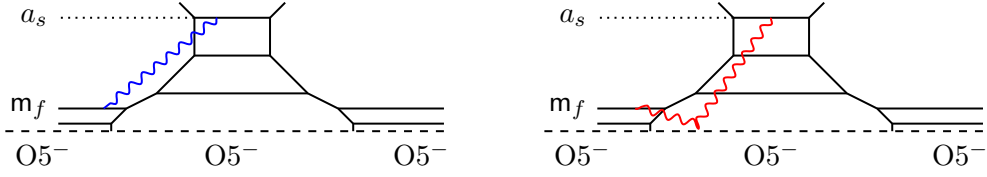
\begin{figure}[H]
    \centering
\begin{minipage}{7cm}
\centering
\begin{tikzpicture}
\draw [dashed] (-1,0)--(4.8,0);
\node at (2,0) [below] {O5${}^-$};
\node at (4.5,0) [below] {O5${}^-$};
\node at (-0.5,0) [below] {O5${}^-$};
\draw (1,0.5)  -- (3,0.5) ;
\draw (1.5,1)  -- (2.5,1);
\draw (1.5,1.5)-- (2.5,1.5);
\draw (0.4,0.1)  --  (-0.3,0.1) ;
\draw (0.6,0.3)  --  (-0.3,0.3) node [left] {$\mathsf{m}_f$};
\draw (3.6,0.1)  -- (4.8,0.1) ;
\draw (3.4,0.3)  -- (4.8,0.3) ;
\draw (0.4,0)--(0.4,0.1)--(0.6,0.3)--(1,0.5)--(1.5,1)--(1.5,1.5)--(1.3,1.7);
\draw (3.6,0)--(3.6,0.1)--(3.4,0.3)--(3,0.5)--(2.5,1)--(2.5,1.5)--(2.7,1.7);
\draw [dotted] (1.5,1.5)-- (-0.3,1.5) node [left] {$a_s$} ;
\draw[decorate,blue, decoration={snake,segment length=6pt, amplitude=1.5pt}] (1.8,1.5) -- (0.3,0.3);
\end{tikzpicture}
\end{minipage}
\begin{minipage}{7cm}
\centering
\begin{tikzpicture}
\draw [dashed] (-1,0)--(4.8,0);
\node at (2,0) [below] {O5${}^-$};
\node at (4.5,0) [below] {O5${}^-$};
\node at (-0.5,0) [below] {O5${}^-$};
\draw (1,0.5)  -- (3,0.5) ;
\draw (1.5,1)  -- (2.5,1);
\draw (1.5,1.5)-- (2.5,1.5);
\draw (0.4,0.1)  --  (-0.3,0.1) ;
\draw (0.6,0.3)  --  (-0.3,0.3) node [left] {$\mathsf{m}_f$};
\draw (3.6,0.1)  -- (4.8,0.1) ;
\draw (3.4,0.3)  -- (4.8,0.3) ;
\draw (0.4,0)--(0.4,0.1)--(0.6,0.3)--(1,0.5)--(1.5,1)--(1.5,1.5)--(1.3,1.7);
\draw (3.6,0)--(3.6,0.1)--(3.4,0.3)--(3,0.5)--(2.5,1)--(2.5,1.5)--(2.7,1.7);
\draw [dotted] (1.5,1.5)-- (-0.3,1.5) node [left] {$a_s$} ;
\draw[decorate,red,decoration={snake,segment length=6pt, amplitude=1.5pt}] (2,1.5) -- (1,0.07) -- (0.2,0.3);
\end{tikzpicture}
\end{minipage}
\caption{Two types of strings that connect the $s$-th color brane and the $f$-th virtual flavor brane. {\sf Left}: the string (in {\color{blue}blue}) connecting them directly. {\sf Right}: the string  (in {\color{red}red}) connecting them through O5-plane. Here, we replaced O5$^+$-planes by the freezing that $\mathrm{O5}^+\sim\mathrm{O5}^-+2\mathrm{D5}$.}
\label{fig:O5-interpret2}
\end{figure}

Consider now the factors in the numerator, 
\begin{align}
\tilde{n}_{\lambda_{s}^T \varnothing} (a_s + \ell {\sf m}_f;\hbar) = \tilde{n}_{\varnothing\lambda_{s} } (a_s + \ell {\sf m}_f;\hbar) \ .
\end{align}
As there are four contributions associated with $\ell m_f$ (with $\ell=\pm 1$), the factor can be interpreted as the contribution from the string connecting the $s$-th color D5-brane associated with Young diagram $\lambda_s$ and the $f$-th flavor D5-brane located at the height $\mathsf{m}_f$ associated with the empty Young diagram. The string connects these two branes directly for $\ell = -1$, while it connects the two branes through the O5-plane for $\ell = +1$. This suggests that there are totally four {\it virtual} flavor branes stuck at the O5-plane: two virtual flavor branes are connected to the NS5-brane on the left and the other two are to the NS5-brane on the right as shown in Figure \ref{fig:O5-interpret2}.

We note that these two virtual flavor branes at each side can be understood from the perspective of the freezing \cite{Hayashi:2023boy, Kim:2024vci}. We can regard an O5$^+$-plane as if it consists of an O5$^-$-plane and 2 D5-branes whose locations are ``frozen,'' and can be symbolically denoted as  O5$^+\sim$ O5$^-+ 2$D5. Such freezing can be generalized to an O$p^+$-plane with generic $p$, 
\begin{align}
\text{``\,O}p^+ \,\sim\, \text{O}p^- \,+\, 2^{p-4}\, \text{D}p\,,\text{''}  
\label{eq:freezing-Op}
\end{align}
where the number of D$p$-branes is the precisely required number for the correct RR-charge.

We also note that one can take a slightly different perspective on the freezing and regard
\begin{equation} 
\boxed{
\begin{aligned}
&\text{``\, O}p^+ \,\sim\, \mathbb{Z}_2 \,+\, 2^{p-5}\, \text{D}p\,,\text{''}\label{eq:freezing-Op-var-1}
\end{aligned}
}
\end{equation}
or 
\begin{align}
&\text{``\,O}p^- \,+\, 2^{p-5} \,\text{D}p\, =:  \text{O}p^0\,\sim\, \mathbb{Z}_2\,,\text{''} 
\label{eq:freezing-Op-var-2}
\end{align}
where $\mathbb{Z}_2$ corresponds to a $\mathbb{Z}_2$ orbifold. 
Such ideas of ``freezing'' or their analog have already appeared, explicitly or implicitly, in the context of cubic prepotentials, Seiberg-Witten curves, Nekrasov partition functions, and superconformal indices \cite{Kim:2021fxx, Kim:2023qwh, Hayashi:2023boy, Kim:2024vci, Lee:2024jae}, 
whose location of the virtual flavor branes is consistent with \eqref{eq:4d-mass}.

\bigskip
\section{SO partition function via 5-brane webs with an O7-plane}\label{sec:SO-via-O7}

Motivated by the interpretation of \eqref{eq:freezing-Op-var-1} and \eqref{eq:freezing-Op-var-2}, discussed in the previous section, we propose a new method for computing topological string partition function for SO gauge theories based on the 5-brane web diagram with O7$^+$-plane and justify our proposal in this section. 

\subsection{5-brane webs with an \texorpdfstring{${\rm O7}^+$}{O7+}-plane}

\begin{figure}[H]
    \centering
\begin{tikzpicture}[scale = 1.1]
\draw [dashed] (-2.5,0)--(2.5,0);
\draw (0,0) circle [radius=0.1] node [below right] {O7${}^+$};
\draw (-1,2)     -- (1,2);
\draw (-0.5,1.5) -- (0.5,1.5);
\draw (-0.5,1)   -- (0.5,1);
\draw (-1,0.5)   -- (1,0.5) ;
\draw (-1,-0.5)  -- (1,-0.5) ;
\draw (-0.5,-1)  -- (0.5,-1);
\draw (-0.5,-1.5)-- (0.5,-1.5);
\draw (-1,-2)    -- (1,-2);
\draw (-2,2.5) -- (-1,2) -- (-0.5,1.5) -- (-0.5,1) -- (-1,0.5) -- (-2,0) 
-- (-1,-0.5) -- (-0.5,-1) -- (-0.5,-1.5) -- (-1,-2) -- (-2,-2.5);
\draw (2,2.5)  --  (1,2) --  (0.5,1.5) --  (0.5,1) --  (1,0.5) -- (2,0) 
-- (1,-0.5)  -- (0.5,-1) --  (0.5,-1.5) --  (1,-2) --  (2,-2.5);
\draw [red, <->, >={Latex[scale=1.2]} ] (-1.33,-0.29) -- (1.33,0.29) ;
\fill (-1.4,-0.3) circle [radius=0.07] ;
\fill (1.4,0.3) circle [radius=0.07] ;
\end{tikzpicture}
\caption{A 5-brane web with an O7${}^+$-plane which describes 5d $\mathcal{N}=1$ pure SO($2N$) gauge theory. The branch cut of an O7${}^+$-plane is depicted by the dashed line. This web diagram includes the covering space which is the part below the branch cut where half of the brane is identified. For instance, consider a point on this web diagram denoted by a thick dot ($\bullet$), there is another point (or dot)  that is identified by $\pi$-rotation with respect to the O7$^+$-plane.}
\label{fig:5-brane-O7}
\end{figure}
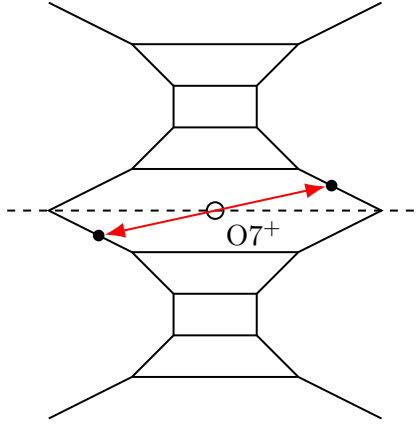

5d $\mathcal{N}=1$ SO($2N$) gauge theories can be realized as a 5-brane web with an orientifold associated with a $\mathbb{Z}_2$ projection in type IIB string theory. One can introduce an O5-plane or an O7$^+$-plane. The former is equipped with a $\mathbb{Z}_2$ (plane-projection) identifying the configuration above the O5-plane with the configuration below, which we have discussed in the previous section. The latter involving an O7$^+$-plane is a $\mathbb{Z}_2$ projection identifying $\pi$-rotated configuration. This is our main interest in this section. 
A representative example of 5d $\mathcal{N}=1$ pure SO($2N$) gauge theory on a 5-brane web is depicted in Figure \ref{fig:5-brane-O7}.

\begin{figure}[H]
    \centering
    \begin{tikzpicture}[x=0.75pt,y=0.75pt,yscale=-0.7,xscale=0.7]

\draw  [dash pattern={on 4.5pt off 4.5pt}]  (382,174) -- (587,173.5) ;
\draw   (573.5,174.28) .. controls (573.56,171.52) and (576.85,169.33) .. (578.61,169.39) .. controls (581.37,169.45) and (583.56,171.74) .. (583.5,174.5) .. controls (583.44,177.26) and (581.15,179.45) .. (578.39,179.39) .. controls (576.63,179.33) and (573.44,177.04) .. (573.5,174.28) -- cycle ;
\draw    (408,173) -- (471,144) ;
\draw    (471,144) -- (575,144) ;
\draw    (471,144) -- (499,113) ;
\draw    (499,113) -- (575,114) ;
\draw    (499,113) -- (500,80) ;
\draw    (500,80) -- (576,81) ;
\draw    (500,80) -- (471,52) ;
\draw    (471,52) -- (575,52) ;
\draw    (408,28) -- (471,52) ;
\draw    (408,173) -- (471,197) ;
\draw    (471,197) -- (575,197) ;
\draw    (500,225) -- (471,197) ;
\draw    (500,225) -- (576,226) ;
\draw    (499,258) -- (500,225) ;
\draw    (499,258) -- (575,259) ;
\draw    (471,289) -- (499,258) ;
\draw    (471,289) -- (575,289) ;
\draw    (408,318) -- (471,289) ;
\draw  [dash pattern={on 4.5pt off 4.5pt}]  (34,221) -- (351,220.5) ;
\draw   (181.5,221.28) .. controls (181.56,218.52) and (183.85,216.33) .. (186.61,216.39) .. controls (189.37,216.45) and (191.56,218.74) .. (191.5,221.5) .. controls (191.44,224.26) and (189.15,226.45) .. (186.39,226.39) .. controls (183.63,226.33) and (181.44,224.04) .. (181.5,221.28) -- cycle ;
\draw    (70,220) -- (133,191) ;
\draw    (133,191) -- (237,191) ;
\draw    (133,191) -- (161,160) ;
\draw    (161,160) -- (209,160.5) ;
\draw    (161,160) -- (162,127) ;
\draw    (162,127) -- (210,127.5) ;
\draw    (162,127) -- (133,99) ;
\draw    (133,99) -- (237,99) ;
\draw    (70,75) -- (133,99) ;
\draw    (237,191) -- (301,221.5) ;
\draw    (237,191) -- (209,160.5) ;
\draw    (209,160.5) -- (210,127.5) ;
\draw    (210,127.5) -- (237,99) ;
\draw    (237,99) -- (300,70) ;

\draw (599,162.4) node [anchor=north west][inner sep=0.75pt]    {$\mathrm{O7}^{+}$};
\draw (173,240.4) node [anchor=north west][inner sep=0.75pt]    {$\mathrm{O7}^{+}$};
\end{tikzpicture}
\caption{Examples of fundamental regions for a 5-brane web with an O7$^+$-plane.}
 \label{fig:fund-region}
\end{figure}
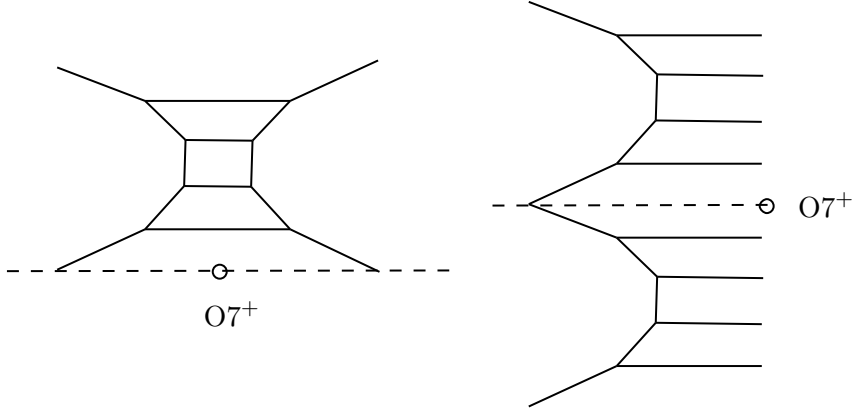

We note that as identifying the $\pi$-rotated part, a 5-brane web with an O7-plane can take various fundamental regions which is the part where the projected part is omitted. Two practically useful fundamental regions are drawn in Figure~\ref{fig:fund-region}.
The one on the left in Figure \ref{fig:fund-region} shows the deficit angle due to an O7$^+$-plane where a $(\pm 2,1)$ 5-brane goes through the cut of the O7$^+$-plane (the dashed line in the figure) reappears as a $(\mp 2,1)$ 5-brane \cite{Bergman:2015dpa}. The one on the right-hand side of Figure \ref{fig:fund-region} shows another fundamental region.

We here propose a new method to compute the topological string partition based on this 5-brane web with an O7$^+$-plane. We observe that the left in Figure \ref{fig:fund-region} is locally identical to the one in Figure \ref{fig:webForSO(2N)} except around the dashed line. Thus, it would be natural to assign the identical factors to the corresponding vertices and edges away from the dashed line. The difference from the computation in section \ref{sec:O-vertex} comes from the contribution coming from the part around the dashed line. In the case with O5$^{\pm}$-plane, we have introduced the O-vertex to deal with this part as we have discussed in section \ref{sec:O-vertex}. In this section, we treat the corresponding part differently.

\begin{figure}[H]
    \centering
    \begin{tikzpicture}[x=0.75pt,y=0.75pt,yscale=-0.7,xscale=0.7]

\draw  [dash pattern={on 4.5pt off 4.5pt}]  (49,193) -- (254,192.5) ;
\draw   (240.5,193.28) .. controls (240.56,190.52) and (242.85,188.33) .. (244.61,188.39) .. controls (248.37,188.45) and (250.56,190.74) .. (250.5,193.5) .. controls (250.44,196.26) and (248.15,198.45) .. (245.39,198.39) .. controls (242.63,198.33) and (240.44,196.04) .. (240.5,193.28) -- cycle ;
\draw    (75,192) -- (138,163) ;
\draw    (138,163) -- (242,163) ;
\draw    (138,163) -- (166,132) ;
\draw    (166,132) -- (242,133) ;
\draw    (166,132) -- (167,99) ;
\draw    (167,99) -- (243,100) ;
\draw    (167,99) -- (138,71) ;
\draw    (138,71) -- (242,71) ;
\draw    (75,47) -- (138,71) ;
\draw    (75,192) -- (138,216) ;
\draw    (138,216) -- (242,216) ;
\draw    (167,244) -- (138,216) ;
\draw    (167,244) -- (243,245) ;
\draw    (166,277) -- (167,244) ;
\draw    (166,277) -- (242,278) ;
\draw    (138,308) -- (166,277) ;
\draw    (138,308) -- (242,308) ;
\draw    (75,337) -- (138,308) ;
\draw    (439,170) -- (502,141) ;
\draw    (502,141) -- (606,141) ;
\draw    (502,141) -- (530,110) ;
\draw    (530,110) -- (606,111) ;
\draw    (530,110) -- (531,77) ;
\draw    (531,77) -- (607,78) ;
\draw    (531,77) -- (502,49) ;
\draw    (502,49) -- (606,49) ;
\draw    (439,25) -- (502,49) ;
\draw    (439,208) -- (502,232) ;
\draw    (502,232) -- (606,232) ;
\draw    (531,260) -- (502,232) ;
\draw    (531,260) -- (607,261) ;
\draw    (530,293) -- (531,260) ;
\draw    (530,293) -- (606,294) ;
\draw    (502,324) -- (530,293) ;
\draw    (502,324) -- (606,324) ;
\draw    (439,353) -- (502,324) ;
\draw [color={rgb, 255:red, 74; green, 144; blue, 226 }  ,draw opacity=1 ][line width=1.5]    (377,169.5) -- (439,170) ;
\draw    (439,170) -- (431,182.5) ;
\draw [color={rgb, 255:red, 74; green, 144; blue, 226 }  ,draw opacity=1 ][line width=1.5]    (378,182.5) -- (431,182.5) ;
\draw [color={rgb, 255:red, 74; green, 144; blue, 226 }  ,draw opacity=1 ][line width=1.5]    (378,194.5) -- (431,194.5) ;
\draw [color={rgb, 255:red, 74; green, 144; blue, 226 }  ,draw opacity=1 ][line width=1.5]    (377,207.5) -- (439,208) ;
\draw    (431,182.5) -- (431,194.5) ;
\draw    (431,194.5) -- (439,208) ;

\draw (260,181.4) node [anchor=north west][inner sep=0.75pt]    {$\mathrm{O7}^{+}$};
\draw (315,185.4) node [anchor=north west][inner sep=0.75pt]  [font=\large]  {$\Longrightarrow $};

\node at (377,169.5) [color={rgb, 255:red, 74; green, 144; blue, 226 },circle,fill,inner sep=1.5pt]{};
\node at (378,182.5) [color={rgb, 255:red, 74; green, 144; blue, 226 },circle,fill,inner sep=1.5pt]{};
\node at (378,194.5) [color={rgb, 255:red, 74; green, 144; blue, 226 },circle,fill,inner sep=1.5pt]{};
\node at (377,207.5) [color={rgb, 255:red, 74; green, 144; blue, 226 },circle,fill,inner sep=1.5pt]{};

\end{tikzpicture}
\caption{An example of 5-brane web with four {\it frozen} D7-branes (denoted by dots in blue).}
 \label{fig:Strip-4D7}
\end{figure}
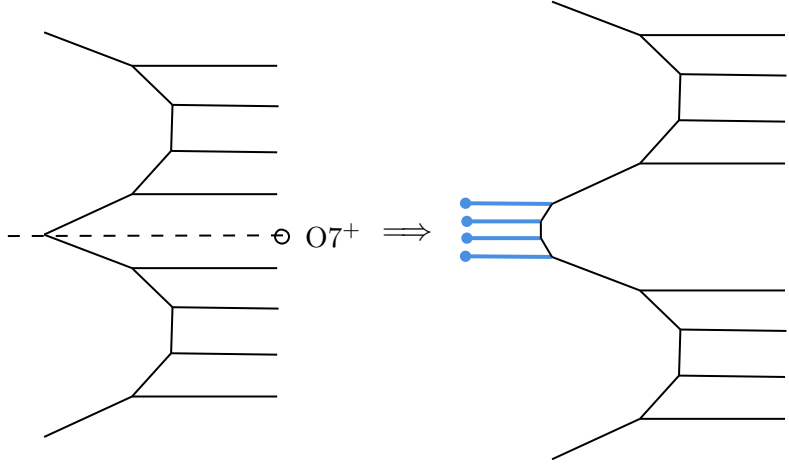

The idea that plays a key role in our proposal 
is to regard the O7$^+$-plane as $\mathbb{Z}_2$-orbifold with four frozen D7-branes:
\begin{align}
\text{``\,O7}^+ \,\sim\, \mathbb{Z}_2\, +\, 4 \,\text{D7\,,''}
\label{eq:O7=Z2+4D7}
\end{align}
which corresponds to the case $p=7$ in \eqref{eq:freezing-Op-var-1}. Using this idea in the right of Figure \ref{fig:fund-region} and pulling out the four D7-branes to the left by the use of the Hanany-Witten transition, we find that the four virtual flavor branes are generated and 
we obtain the strip diagram in the right of Figure \ref{fig:Strip-4D7}.

Due to this process, we can now treat this fundamental region as a single strip. Furthermore, we identify these four virtual flavor branes as the ones discussed in section~\ref{sec:int-freeze}. This indicates that the heights of these four branes are given by~\eqref{eq:4d-mass}.

\subsection{Topological string partition function}\label{sebsec:top-string}

In this subsection, we compute the topological string partition function for 5d $\mathcal{N}=1$ pure SO($2N$) gauge theories based on the strip 5-brane web diagram given in the right of Figure~\ref{fig:Strip-4D7}.

\begin{figure}[H]
\centering
\begin{tikzpicture}[x=0.75pt,y=0.75pt,yscale=-0.7,xscale=0.7]
\draw    (270,20) -- (330,45) -- (365,80) -- (365,115) -- (330,150)  
-- (255,180) -- (240,195) -- (240, 210) -- (255, 225) -- (330,250)
-- (365, 285) -- (365,320) -- (330,355) -- (270,380);
\draw [color={rgb, 255:red, 74; green, 144; blue, 226 }  ,draw opacity=1 ][line width=1.5]    (180,180) -- (255,180) ;
\draw [color={rgb, 255:red, 74; green, 144; blue, 226 }  ,draw opacity=1 ][line width=1.5]    (180,195) -- (240,195) ;
\draw [color={rgb, 255:red, 74; green, 144; blue, 226 }  ,draw opacity=1 ][line width=1.5]    (180,210) -- (240,210) ;
\draw [color={rgb, 255:red, 74; green, 144; blue, 226 }  ,draw opacity=1 ][line width=1.5]    (180,225) -- (255,225) ;
\node at (180,180) [color={rgb, 255:red, 74; green, 144; blue, 226 } ,circle,fill,inner sep=1.5pt]{};
\node at (180,195) [color={rgb, 255:red, 74; green, 144; blue, 226 } ,circle,fill,inner sep=1.5pt]{};
\node at (180,210) [color={rgb, 255:red, 74; green, 144; blue, 226 } ,circle,fill,inner sep=1.5pt]{};
\node at (180,225) [color={rgb, 255:red, 74; green, 144; blue, 226 } ,circle,fill,inner sep=1.5pt]{};
\draw  (330,45) -- (400,45) ;
\draw    (400,45) -- (455,45) ;
\draw    (365,80) -- (455,80) ;
\draw    (365,115) -- (455,115) ;
\draw    (330,150) -- (455,150) ;
\draw    (330,250) -- (455,250) ;
\draw    (365,285) -- (455,285) ;
\draw    (365,320) -- (455,320) ;
\draw    (330,355) -- (455,355) ;
\draw [->, >={Latex[scale=1.2]}] (400,45)--(405,45) ;
\draw [->, >={Latex[scale=1.2]}] (400,80)--(405,80) ;
\draw [->, >={Latex[scale=1.2]}] (400,115)--(405,115) ;
\draw [->, >={Latex[scale=1.2]}] (400,150)--(405,150) ;
\draw [->, >={Latex[scale=1.2]}] (400,250)--(405,250) ;
\draw [->, >={Latex[scale=1.2]}] (400,285)--(405,285) ;
\draw [->, >={Latex[scale=1.2]}] (400,320)--(405,320) ;
\draw [->, >={Latex[scale=1.2]}] (400,355)--(405,355) ;
\draw (400,45)  node [anchor=south west][inner sep=1.5pt]    {$\lambda_N$};
\draw (400,115) node [anchor=south west][inner sep=1.5pt]    {$\lambda_2$};
\draw (400,150) node [anchor=south west][inner sep=1.5pt]    {$\lambda_1$};
\draw (400,250) node [anchor=south west][inner sep=1.5pt]    {$\lambda_1{}^{T}$};
\draw (400,285) node [anchor=south west][inner sep=1.5pt]    {$\lambda_2{}^{T}$};
\draw (400,355) node [anchor=south west][inner sep=1.5pt]    {$\lambda_N{}^{T}$};
\draw (460,45)  node [anchor=west]  {$A_{N}$};
\draw (465,75)  node [anchor=west]  {$\vdots$};
\draw (460,115) node [anchor=west]  {$A_{2}$};
\draw (460,150) node [anchor=west]  {$A_{1}$};
\draw (460,250) node [anchor=west]  {$A_{1}^{-1}$};
\draw (460,280) node [anchor=west]  {$A_{2}^{-1}$};
\draw (465,315) node [anchor=west]  {$\vdots$};
\draw (460,355) node [anchor=west]  {$A_{N}^{-1}$};
\draw (30,257) node [anchor=west] {${\sf M}_{f} =\big\{1, -1, q^{1/2},  -q^{1/2}\big\}$};

\end{tikzpicture}
\caption{The parametrization for the strip diagram. The $\mathbb{Z}_2$ orbifold action ensures $A^{-1}_n$ and $\lambda^T_n$ for the lower part of the strip and four frozen D7-branes are denoted by four dots (in blue) whose positions are denoted by $\mathsf{M}_f$. }  
\label{fig:stripdiagram2}
\end{figure}
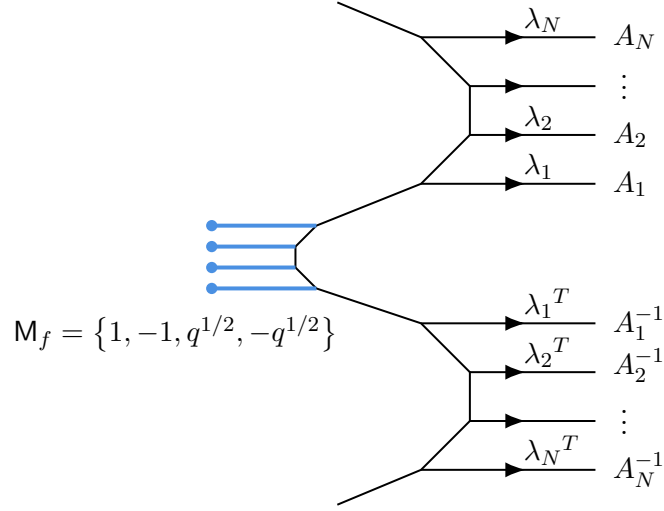

We denote the heights of the color branes relative to the O7$^+$-plane by $a_{N}, \cdots a_{1}$ respectively, from the top. This parameterization is essentially identical to the ones in Figure \ref{fig:webForSO(2N)}. Due to the $\mathbb{Z}_2$ orbifold action, the heights of the remaining $N$ color branes should be $-a_{N}, \cdots, -a_{1}$ from the bottom.  Or, if we use the variables $A_s = e^{-a_s}$, the location of the color branes are given by $A_N, \cdots, A_1, A_1^{-1}, \cdots, A_N^{-1}$ from the top as given in Figure \ref{fig:stripdiagram2}.  Also, as mentioned in the previous subsection, we assume that the heights of the four virtual flavor branes are given by $\mathsf{m}_f$ ($f=1,2,3,4$) in \eqref{eq:4d-mass} or $\mathsf{M}_f = e^{-\mathsf{m}_f}$ in \eqref{eq:defM1234}.
 
Suppose that we assign Young diagrams $\lambda_{s}$ to the color brane at the height $a_{s}$, which assignment is again identical to the one in Figure \ref{fig:webForSO(2N)}. Since the color brane at the height $a_s$ is identified with the color brane at the height $-a_{s}$, we should assign the same Young diagram. 
However, the assignment of the arrows for the identified color branes are opposite to each other, for instance, the direction of the arrow for the color brane at the height $-a_s$ becomes the opposite of that for the color brane at $a_s$, 
so that the direction of the arrow is consistent with the $\mathbb{Z}_2$ orbifold action. 
Taking into account that changing the direction of the arrow is equivalent to transposing the Young diagram, we assign $\lambda_s^T$ to the color brane with the height $-a_s$ and, at the same time, we change the direction of the arrow assigned to it as in Figure \ref{fig:stripdiagram2}. 

\begin{figure}[H]
    \centering
    \begin{tikzpicture}[x=0.75pt,y=0.75pt,yscale=-0.6,xscale=0.6]
\draw    (300,350) -- (450,350) ;
\draw [->, >={Latex[scale=1.0]}] (300,350)--(390,350) ;
\draw    (300,250) -- (450,250) ;
\draw [->, >={Latex[scale=1.0]}] (300,250)--(390,250) ;
\draw    (300,150) -- (450,150) ;
\draw [->, >={Latex[scale=1.0]}] (300,150)--(390,150) ;
\draw    (300,300) -- (150,300) ;
\draw [->, >={Latex[scale=1.0]}] (150,300)--(240,300) ;
\draw    (300,200) -- (150,200) ;
\draw [->, >={Latex[scale=1.0]}] (150,200)--(240,200) ;
\draw    (300,100) -- (150,100) ;
\draw [->, >={Latex[scale=1.0]}] (150,100)--(240,100) ;
\draw    (300,30) -- (300,370) ;
\draw (370,50) node [anchor=north west][inner sep=0.75pt]    {$\vdots$};
\draw (470,140) node [anchor=north west][inner sep=0.75pt]    {$B_{3}$};
\draw (470,240) node [anchor=north west][inner sep=0.75pt]    {$B_{2}$};
\draw (470,340) node [anchor=north west][inner sep=0.75pt]    {$B_{1}$};

\draw (220,30) node [anchor=north west][inner sep=0.75pt]    {$\vdots$};
\draw (110,290) node [anchor=north west][inner sep=0.75pt]    {$A_{1}$};
\draw (110,190) node [anchor=north west][inner sep=0.75pt]    {$A_{2}$};
\draw (110,90) node [anchor=north west][inner sep=0.75pt]    {$A_{3}$};
\draw (370,120) node [anchor=north west][inner sep=0.75pt]    {$\nu_{3}$};
\draw (370,220) node [anchor=north west][inner sep=0.75pt]    {$\nu_{2}$};
\draw (370,320) node [anchor=north west][inner sep=0.75pt]    {$\nu_{1}$};

\draw (220,270) node [anchor=north west][inner sep=0.75pt]    {$\mu_{1}$};
\draw (220,170) node [anchor=north west][inner sep=0.75pt]    {$\mu_{2}$};
\draw (220,70) node [anchor=north west][inner sep=0.75pt]    {$\mu_{3}$};

\end{tikzpicture}
\caption{A strip diagram. Here, the charges of 5-branes are neglected. 
$A_n, \mu_n$ are K\"ahler parameters and the corresponding Young diagrams for D5-branes on the left, while $B_n, \nu_n$ are K\"ahler parameters and the corresponding Young diagrams for D5-branes on the right.
}
 \label{fig:stripdiagram}
\end{figure}
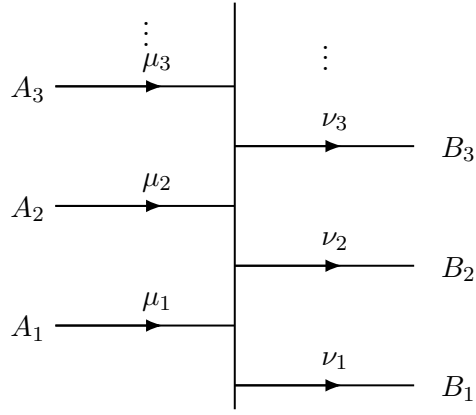

We now compute the amplitude of this strip diagram. In general, the strip diagram is given as in Figure \ref{fig:stripdiagram}, where we denote all the $(p,1)$ 5-branes as the vertical lines for simplicity. We denote the height of the horizontal lines on the left as $A_k$ while the one on the right as $B_s$. We assign the Young diagram $\mu_k$ to the line at the height $A_k$ on the left and $\nu_k$ to the line at the height $B_s$ on the right.

Following the computation given by \cite{Iqbal:2004ne}, we find that the amplitude for the strip diagram in Figure \ref{fig:stripdiagram} is given by \cite{Li:2021rqr}
\begin{align}
Z_{\text{strip}} =
& \prod_{k} s_{\mu_k}(q^{-\rho})
\times 
\prod_{s} s_{\nu_s^T}(q^{-\rho})
\\
&\times
\prod_{k<l} X^{-1}_{\mu_k, \mu_l} ( A_k, A_l ) 
\times 
\prod_{s<t} X^{-1}_{\nu_s, \nu_t} ( B_s, B_t )
\times
\prod_{k, s} X_{\mu_k, \nu_s}  \left( A_k,  B_s \right)\nonumber
\end{align}
with
\begin{align}
X_{\mu , \nu } \left( A,  B  \right)
:= \left\{
\begin{array}{ll}
R_{\nu{}^T \mu}\left( A B^{-1} \right) & \quad \text{ if }  \,\, |A| < |B|
\\
R_{ \mu{}^T \nu} \left( A^{-1} B \right) & \quad \text{ if } \,\, |B| < |A|
\end{array}
\right. .
\end{align}

Applying this formula to the diagram in Figure~\ref{fig:stripdiagram2} equipped with $\mu_f=\varnothing$, 
assuming $|A_N| < \cdots < |A_1| < |\mathsf{M}_4| = |\mathsf{M}_3| = q^\frac12< |\mathsf{M}_2| = |\mathsf{M}_1| = 1$, 
we find that this strip amplitude is given by%
\footnote{
Since $|\mathsf{M}_1| = |\mathsf{M}_2|$ and $|\mathsf{M}_3| = |\mathsf{M}_4|$, there are some issues in applying this formula. The prescription we use here is to deform $\mathsf{M}_2 \to e^{-\varepsilon} \mathsf{M}_2$, $\mathsf{M}_4 \to e^{-\varepsilon} \mathsf{M}_4$ with $\varepsilon > 0$ and then to take the limit $\varepsilon \to 0$ after using this formula. Although a different prescription may give slightly different results, it affects only the extra factor, which we finally discard. Thus, this problem is not very crucial in this paper.
}
\begin{align}\label{strip-SO2N}
Z_{\text{strip}} =
& 
\prod_{s=1}^N s_{\lambda_s }(q^{-\rho}) s_{\lambda_s^T}(q^{-\rho})
\prod_{1 \le k < \ell \le 4} R^{-1}_{\varnothing\varnothing}({\sf M}_k^{-1} {\sf M}_\ell)
\cr
&\times \prod_{1\le s<t \le N}  R^{-2}_{\lambda_{s}^T \lambda_{t}} (A_s^{-1} A_t) 
\times
\prod_{s=1}^N \prod_{t=1}^N R^{-1}_{\lambda_{s} \lambda_{t}} (A_s A_t) 
\cr
&\times 
\prod_{s=1}^N \prod_{f=1}^4 \prod_{\ell =\pm 1} R_{\lambda_{s} \varnothing}({\sf M}_f^{\ell}A_s )\ .
\end{align}

In order to obtain the topological string partition function, we need to include the contribution from the horizontal edges and sum over the Young diagrams associated with them. For this purpose, it would be better to use the left in Figure \ref{fig:fund-region}. Since the structures around the horizontal lines are locally the same as those in Figure \ref{fig:webForSO(2N)}, we can glue the strip diagram analogous to \eqref{eq:glueLR} as
\begin{align}
Z&=\sum_{\vec{\lambda}} Z_{\text{strip}}
\prod_{s=1}^N (-Q_{B s})^{|\lambda_s|} f_{\lambda_s}^{5-2s},
\end{align}
where $Q_{Bs}$ ($s=1,\cdots, N$) are given in \eqref{eq:ParamQB}.

Finally, in order to obtain the topological string partition function that agrees with the Nekrasov partition function, we need to remove the ``extra factor'' as has been discussed in various works including \cite{Bergman:2013ala, Bergman:2013aca, Bao:2013pwa, Hayashi:2013qwa, Hwang:2014uwa}. The extra factor, which does not include the Coulomb moduli dependence, can be extracted by taking the limit such that all the color branes including their mirror images are separated enough from each other. That is, $|A_{s+1}/A_s| \ll 1$ for $s=1,2,\cdots, N-1$ and $|A_1| \ll 1$, which indicates $|Q_{Bs}| \ll 1$, $|A_s^{-1} A_t| \ll 1$ for $s<t$, and $|A_s A_t| \ll 1$. In this limit, we find that  
\begin{align}
Z \to \prod_{1 \le k < \ell \le 4} R^{-1}_{\varnothing\varnothing}({\sf M}_k^{-1} {\sf M}_\ell) =: Z_{\text{extra}},
\end{align}
which we identify as an extra factor. Thus, our topological string partition function for 5d $\mathcal{N}=1$ SO($2N$) gauge theory computed based on the 5-brane web with O7$^+$-plane is given by
\begin{align}\label{eq:top-par-O7}
Z_{{\rm SO}(2N)}^{\textrm{top}} 
= \frac{Z}{Z_{\textrm{extra}}} 
=& \sum_{\vec{\lambda}} 
\prod_{s=1}^N (-Q_{B s})^{|\lambda_s|} f_{\lambda_s}^{5-2s}
\prod_{s=1}^N s_{\lambda_s }(q^{-\rho}) s_{\lambda_s^T}(q^{-\rho})
\cr
&\qquad \times 
\prod_{1\le s<t \le N}  R^{-2}_{\lambda_{s}^T \lambda_{t}} (A_s^{-1} A_t) 
\times
\prod_{s=1}^N \prod_{t=1}^N R^{-1}_{\lambda_{s} \lambda_{t}} (A_s A_t)
\cr
&\qquad \times 
\prod_{r=1}^N \prod_{f=1}^4 \prod_{\ell =\pm 1} R_{\lambda_{r} \varnothing}({\sf M}_f^{\ell}A_{r} )\ .
\end{align}

\bigskip

\subsection{Consistency check and the agreement}

The partition function \eqref{eq:top-par-O7} is obtained using our new method based on the 5-brane web with O7$^+$-plane as a $\mathbb{Z}_2$ projection equipped with four virtual flavor branes. Although we have proposed this method motivated by the computation in section \ref{sec:SO-via-O5}, it would be still meaningful to check explicitly that \eqref{eq:top-par-O7} exactly reproduces the same result obtained from the 5-brane web with O5-plane.


We begin with the perturbative part by taking limit $Q_I \to 0$ in \eqref{eq:top-par-O7} as done in  \eqref{eq:top-Qto0}: 
\begin{align}\label{eq:QI0-O7}
\lim_{Q_I \to 0} Z_{{\rm SO}(2N)}^{\textrm{top}} 
= 
&
\prod_{1 \le s<t \le N} \mathrm{PE} \lt( \frac{2q}{(1-q)^2} A_s^{-1} A_t \rt)
\times
\prod_{s=1}^N \prod_{t=1}^N \mathrm{PE} \lt( \frac{q}{(1-q)^2} A_s A_t \rt)
\cr
&\qquad \times 
\prod_{s=1}^N \prod_{f=1}^4 \prod_{\ell =\pm 1} 
\mathrm{PE} \lt( - \frac{q}{(1-q)^2} {\sf M}_f^{\ell}A_{s} \rt),
\end{align}
where we have used \eqref{eq:def-R-N}.
As each factor in \eqref{eq:QI0-O7} is re-expressed respectively as 
\begin{align}
\prod_{1 \le s<t \le N} \mathrm{PE} \lt( \frac{2q}{(1-q)^2} A_s^{-1} A_t \rt)
&=  \mathrm{PE} \lt( \frac{2q}{(1-q)^2} \sum_{1 \le s<t \le N}\!\! A_s^{-1} A_t \rt),
\cr
\prod_{s=1}^N \prod_{t=1}^N \mathrm{PE} \lt( \frac{q}{(1-q)^2} A_s A_t \rt)
&=  \mathrm{PE} \lt( \frac{q}{(1-q)^2} \Bigg( 2 \!\!\!\sum_{1 \le s<t \le N}\!\!\! A_s A_t + \sum_{s=1}^N A_s^2 \Bigg)\rt),
\cr
\prod_{s=1}^N \prod_{f=1}^4 \prod_{\ell =\pm 1} 
\mathrm{PE} \lt( -\frac{q}{(1-q)^2} {\sf M}_f^{\ell}A_s \rt)
&=  \mathrm{PE} \lt( -\sum_{s=1}^N \frac{q}{(1-q)^2} A_s^{2} \rt)\ ,
\end{align}
the perturbative part \eqref{eq:QI0-O7} amounts to the expected perturbative part:
\begin{align}\label{eq:pert-from-O7}
\lim_{Q_I \to 0} Z_{{\rm SO}(2N)}^{\textrm{top}} 
= 
\mathrm{PE} \lt( \frac{2q}{(1-q)^2}\!\! \sum_{1 \le s<t \le N}\!\!\! (A_s^{-1} A_t + A_s A_t) \rt)\ ,
\end{align}
which is the same as \eqref{pert-part} obtained from a 5-brane web with an O5-plane.

The instanton part is computed by dividing the topological string partition function \eqref{eq:top-par-O7} by the perturbative part \eqref{eq:QI0-O7}:
\begin{align}\label{}
Z_{{\rm SO}(2N)}^{\textrm{inst}}
&= \frac{Z_{{\rm SO}(2N)}^{\textrm{top}}}{\displaystyle\lim_{Q_I \to 0} Z_{{\rm SO}(2N)}^{\textrm{top}}}
\cr
&=\sum_{\vec{\lambda}} \prod_{s=1}^N (-Q_{B s})^{|\lambda_{s}|} f_{\lambda_{s}}^{5-2s} 
s_{\lambda_s}(q^{-\rho}) s_{\lambda_s^T}(q^{-\rho})
\times  \!\!
\prod_{1\le s< t\le N}  \!\!N^{-2}_{\lambda_{s}\lambda_{t}}(A_{t}A_s^{-1};q)
\cr
 & \quad \times 
\prod_{s=1}^N \prod_{t=1}^N 
N^{-1}_{\lambda_{s}^T \lambda_t}(A_s A_t;q)
\times \prod_{s=1}^N \prod_{f=1}^4 \prod_{\ell=\pm 1}
N_{\lambda_{s}^T \varnothing} (A_s {\sf M}_f^{\ell} ;q)\ ,
\end{align}
where we used \eqref{eq:def-R-N}.
This indeed agrees with the instanton part \eqref{eq:SO-inst-1}, computed based on the 5-brane web with an O5-plane. We note that this agreement on the instanton partition function means that the expression \eqref{eq:5D-SO-final} for the Nekrasov instanton partition function for 5d $\mathcal{N}=1$ pure SO($2N$) gauge theory can also be derived by our topological vertex formalism with an O7$^+$-plane.

\subsection{Comments on the folding}\label{sec:folding}

In \cite{Lee:2024jae}, it was reported that  
an intriguing relation between the SO($2N$) partition function and the folding of the SU($2N$) partition function based on the ADHM construction. It is worth noting that our result is also consistent with this relation.

An intuitive way to see this relation is as follows.
Observe that the strip diagram in Figure \ref{fig:stripdiagram2} looks like a `half'' of a 5-brane web diagram for SU($2N$) gauge theory with the restriction that the Coulomb branch parameters, flavor masses, and the Young diagrams assigned to the color branes are tuned in such a specific way that respects a $\mathbb{Z}_2$ projection. In other words, the strip diagram in Figure \ref{fig:stripdiagram2} looks like the $\mathbb{Z}_2$ folding of the SU($2N$) gauge theory with eight flavors.

\begin{figure}[H]
\centering
\begin{tikzpicture}[x=0.75pt,y=0.75pt,yscale=-0.7,xscale=0.7]
\draw    (270,20) -- (330,45) -- (365,80) -- (365,115) -- (330,150)  
-- (255,180) -- (240,195) -- (240, 210) -- (255, 225) -- (330,250)
-- (365, 285) -- (365,320) -- (330,355) -- (270,380);
\draw [color={rgb, 255:red, 74; green, 144; blue, 226 },draw opacity=1 ][line width=1]   
(180,180) -- (255,180) ;
\draw [color={rgb, 255:red, 74; green, 144; blue, 226 },draw opacity=1 ][line width=1]    (180,195) -- (240,195) ;
\draw [color={rgb, 255:red, 74; green, 144; blue, 226 },draw opacity=1 ][line width=1]    (180,210) -- (240,210) ;
\draw [color={rgb, 255:red, 74; green, 144; blue, 226 },draw opacity=1 ][line width=1]    (180,225) -- (255,225) ;

\draw  (330,45) -- (400,45) ;
\draw    (400,45) -- (455,45) ;
\draw    (365,80) -- (455,80) ;
\draw    (365,115) -- (455,115) ;
\draw    (330,150) -- (455,150) ;
\draw    (330,250) -- (455,250) ;
\draw    (365,285) -- (455,285) ;
\draw    (365,320) -- (455,320) ;
\draw    (330,355) -- (455,355) ;
\node at (180,180) [color={rgb, 255:red, 74; green, 144; blue, 226 },circle,fill,inner sep=1.5pt]{};
\node at (180,195) [color={rgb, 255:red, 74; green, 144; blue, 226 }, circle,fill,inner sep=1.5pt]{};
\node at (180,210) [color={rgb, 255:red, 74; green, 144; blue, 226 }, circle,fill,inner sep=1.5pt]{};
\node at (180,225) [color={rgb, 255:red, 74; green, 144; blue, 226 },circle,fill,inner sep=1.5pt]{};

\draw [->, >={Latex[scale=1.2]}] (400,45)--(405,45) ;
\draw [->, >={Latex[scale=1.2]}] (400,80)--(405,80) ;
\draw [->, >={Latex[scale=1.2]}] (400,115)--(405,115) ;
\draw [->, >={Latex[scale=1.2]}] (400,150)--(405,150) ;
\draw [->, >={Latex[scale=1.2]}] (400,250)--(405,250) ;
\draw [->, >={Latex[scale=1.2]}] (400,285)--(405,285) ;
\draw [->, >={Latex[scale=1.2]}] (400,320)--(405,320) ;
\draw [->, >={Latex[scale=1.2]}] (400,355)--(405,355) ;
\draw (400,45)  node [anchor=south west][inner sep=1.5pt]    {$\lambda_N$};
\draw (400,115) node [anchor=south west][inner sep=1.5pt]    {$\lambda_2$};
\draw (400,150) node [anchor=south west][inner sep=1.5pt]    {$\lambda_1$};
\draw (400,250) node [anchor=south west][inner sep=1.5pt]    {$\lambda_{-1}$};
\draw (400,285) node [anchor=south west][inner sep=1.5pt]    {$\lambda_{-2}$};
\draw (400,355) node [anchor=south west][inner sep=1.5pt]    {$\lambda_{-N}$};
\draw (460,45)  node [anchor=west]  {$a_N$ ($A_{N}=e^{-a_N})$};
\draw (465,75)  node [anchor=west]  {$\vdots$};
\draw (460,115) node [anchor=west]  {$a_{2}$};
\draw (460,150) node [anchor=west]  {$a_{1}$};
\draw (460,250) node [anchor=west]  {$a_{-1}$};
\draw (460,280) node [anchor=west]  {$a_{-2}$};
\draw (465,315) node [anchor=west]  {$\vdots$};
\draw (460,355) node [anchor=west]  {$a_{-N}$};

\draw (30,258) node [anchor=west] {${\sf m}_{f} = \{0, \pi i, \frac{\hbar}{2}, \frac{\hbar}{2}+\pi i\}$};
\draw (190,169) node [anchor=west] {$\varnothing$};

\end{tikzpicture}
\caption{New parametrization for the strip diagram: $a_{-i}:=-a_i$ and $\lambda_{-i}:=\lambda_i^T$.} 
\label{fig:stripdiagram2-new}
\end{figure}
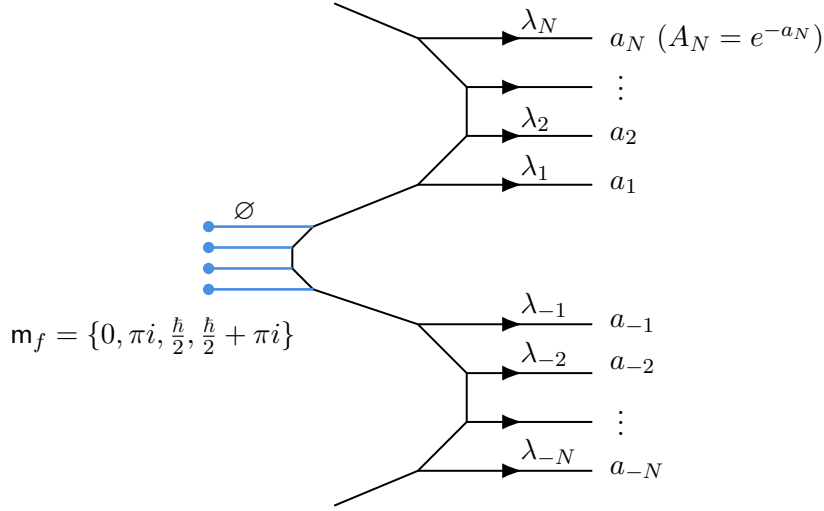

In the following, we see this observation more quantitatively. Suppose we assign labels $s=N,N-1,\cdots,1,-1,,\cdots, -N$ to the color branes from the above in Figure \ref{fig:stripdiagram2}. We denote the height of the color brane labeled by $s$ as $a_s$ and assign Young diagram $\lambda_s$ to it. Then, we can make a new labeling, made in Figure \ref{fig:stripdiagram2}, to the following: 
\begin{align}\label{eq:folding-SU=tune}
a_{-s} := - a_{s},
\quad
\lambda_{-s} := \lambda_{s}^T,
\qquad
(s=1,2,\cdots, N)\ , 
\end{align}
which is depicted in Figure \ref{fig:stripdiagram2-new}.

Employing the properties of new Nekrasov factor \eqref{eq:newNekr-rel}, one immediately finds that the Nekrasov instanton partition function \eqref{eq:5D-SO-final} for 5d $\mathcal{N}=1$ pure SO($2N$) gauge theory can be rewritten as 
\begin{align}
    Z_{{\rm SO}(2N)}^{\textrm{inst}}
=& \sum_{\vec{\lambda}} Q_I^{|\vec{\lambda}|} 
\frac{\displaystyle
\prod_{\substack{-N \le s \le N \\ s \neq 0}}\prod_{f=1}^4  
\tilde{n}_{\varnothing \lambda_{s}} (a_s - {\sf m}_f;\hbar)
}{\displaystyle
\prod_{\substack{-N \le s \le N \\ s \neq 0}} \prod_{t=1}^N 
\tilde{n}_{\lambda_{s} \lambda_{t}}(a_t-a_s;\hbar) }\ . 
\end{align}

Also, in addition to the four masses ${\sf m}_{f}$ given in \eqref{eq:4d-mass}, we introduce four more masses ${\sf m}_{-f}$ defined by
\begin{align}\label{eq:folding-mass}
&{\sf m}_{-f} := - {\sf m}_{f},
\qquad 
(f=1,2,3,4).
\end{align}

With these notations, we find that we can rewrite the Nekrasov instanton partition function \eqref{eq:5D-SO-final} for 5d $\mathcal{N}=1$ pure SO($2N$) gauge theory as 
\begin{align}
    Z_{{\rm SO}(2N)}^{\textrm{inst}}
=& 
\sum_{\vec{\lambda}} Q_I^{|\vec{\lambda}|} 
\left(
\frac{\displaystyle
\prod_{\substack{-4 \le f \le 4 \\ f \neq 0}} 
\prod_{\substack{-N \le t \le N \\ t \neq 0}}
\tilde{n}_{\varnothing \lambda_{t}} (a_t - {\sf m}_f;\hbar)
}{\displaystyle
\prod_{\substack{-N \le s \le N \\ s \neq 0}}
 \prod_{\substack{-N \le t \le N \\ t \neq 0}}
\tilde{n}_{\lambda_{s} \lambda_{t}}(a_t-a_s;\hbar) }
\right)^{\frac12}.
\end{align}
We observe that the summand is identical to the square root of the summand of the partition function for the SU($2N$) gauge theory with eight flavors under the tuning \eqref{eq:folding-SU=tune} and \eqref{eq:folding-mass}. This observation is consistent with what is discussed in \cite{Lee:2024jae}.

We can make analogous observations also for SO($2N+1$) gauge theory. In addition to \eqref{eq:folding-SU=tune} and \eqref{eq:folding-mass}, we define 
\begin{align}\label{eq:folding-SUodd-tune}
&{\sf m}_{0} := 0,
\qquad
a_{0} := 0,
\qquad
\lambda_{0} := \varnothing.
\end{align}
Then, we can rewrite \eqref{eq:5D-SOodd-final} as
\begin{align}
    Z_{{\rm SO}(2N+1)}^{\textrm{inst}}
&= \sum_{\vec{\lambda}} Q_I^{|\vec{\lambda}|} 
\frac{\displaystyle
\prod_{-N \le s \le N} \prod_{f=1}^4 
\tilde{n}_{\varnothing \lambda_{s}} (a_s - {\sf m}_f;\hbar)
}{\displaystyle
\prod_{-N \le s \le N} \prod_{t=1}^N 
\tilde{n}_{\lambda_{s} \lambda_{t}}(a_t-a_s;\hbar) }
\cr
&= \sum_{\vec{\lambda}} Q_I^{|\vec{\lambda}|} 
\lt( \frac{\displaystyle
\prod_{-4 \le f \le 4} \prod_{-N \le t \le N}
\tilde{n}_{\varnothing \lambda_{t}} (a_t - {\sf m}_f;\hbar)
}{\displaystyle
\prod_{-N \le s \le N} \prod_{-N \le t \le N} 
\tilde{n}_{\lambda_{s} \lambda_{t}}(a_t-a_s;\hbar) }
\rt)^{\frac12}.
\end{align}
We observe that the summands are the square roots of the ones for the SU$(2N+1)$ gauge theory with nine flavors.

\bigskip
\section{Conclusion}\label{sec:conclusion}

In this paper, we computed the topological string partition function corresponding to the Nekrasov instanton partition function for 5d $\mathcal{N}=1$ pure SO($2N$) gauge theory compactified on an S$^1$. The corresponding 5-brane web can be constructed with two distinct orientifolds: one with an O5-plane and the other with an O7$^+$-plane. We applied the topological vertex formalism to compute the partition function based on these two orientifolds as follows. For the case with an O5-plane, we implemented the O-vertex proposed in \cite{Hayashi:2020hhb}. The resulting partition function is expressed in terms of the M-factor as given in \eqref{eq:SO-via-M}. We showed that the M-factor is rewritten as a product of the Nekrasov factors. A redefinition of the Nekrasov factor allows us to give an intuitive interpretation of the instanton partition functions as a product of the contributions from W-bosons of SO($2N$) gauge theory and also the contributions from four virtual flavors. Our interpretation is also consistent with the freezing of O5$^+$-plane proposed in \cite{Hayashi:2023boy, Kim:2024vci} that $\mathrm{O5}^+\sim\mathrm{O5}^-+2\mathrm{D5}$, which also suggests that the charge neutral combination $\mathrm{O5}^0=\mathrm{O5}^-+\mathrm{D5}$ \cite{Sen:1998rg,Sen:1998ii,Hanany:1999sj} may serve as an $\mathbb{Z}_2$ orbifold.

Motivated by this interpretation that the neutral combination $\mathrm{Op}^-+2^{p-5}\mathrm{Dp}$ can be regarded as a $\mathbb{Z}_2$ orbifold, we sought to apply it to a 5-brane web with an O7$^{+}$-plane describing SO($2N$) gauge theory and proposed a topological vertex formalism with an O7$^{+}$-plane. Our proposal is that on the 5-brane web, an O7$^{+}$-plane is substituted with the $\mathbb{Z}_2$ orbifold accompanied with four (frozen) D7-branes. We confirmed that the topological string partition function computed based on this proposal coincides with the one computed from the 5-brane web with an O5-plane. We also showed that our result is consistent with \cite{Lee:2024jae}, which studies the relation between SO($2N$) and the folding of SU($2N$) based on the ADHM construction.

It is worth noting that our proposal substituting an O7$^{+}$-plane with the $\mathbb{Z}_2$ orbifold and four (frozen) D7-branes may also be interpreted as a generalization of the freezing discussed in \cite{Hayashi:2023boy}, which computes the Seiberg-Witten curve for the SO($2N$) gauge theory based on a 5-brane web diagram with an O7$^+$-plane. Indeed, the tuning \eqref{eq:defM1234} or \eqref{eq:4d-mass} is nothing but the freezing constraint imposed on the Seiberg-Witten curve in the vanishing limit of the Omega deformation parameter $\hbar \to 0$.

To emphasize our proposal, we have focused on the computations of SO($2N$) gauge theory without any hypermultiplets, but it is straightforward to apply it to the case with hypermultiplets in the fundamental representation and also to SO($2N$)-SU($N'$) quiver gauge theories which can be realized with an O7$^+$-plane.

One interesting direction to pursue is to check whether one can generalize our method to  5-brane web diagrams with an O7$^-$-plane without resolving an O7$^-$-plane into two 7-branes \cite{Sen:1996vd}, which includes the one corresponding to 5d $\mathcal{N}=1$ Sp($N$) gauge theory. As reported in \cite{Nawata:2021dlk}, the Nekrasov instanton partition function for the Sp($N$) gauge theory contains four additional Young diagram sums with four extra Coulomb moduli parameters with a similar tuning to \eqref{eq:defM1234}. It would be interesting to further develop our proposal to ascertain whether there could be some hidden connection to the additional Young diagram sums presented in \cite{Nawata:2021dlk}. It would also be beneficial to extend our proposal to encompass the refined topological vertex.

\acknowledgments
We thank Kimyeong Lee, Song Liu, Satoshi Nawata, Ashoke Sen, and Xin Wang for useful discussions. We thank Fudan university for hosting ``String Theory and Quantum Field Theory 2024'' and UESTC and Sichuan university for hosting ``Tianfu Fields and Strings 2024,''and Soochow university for hosting ``2nd High-Energy symposium'' where part of the work is done. SK also thanks Tianjin university, Xiamen university, and SIMIS for the hospitality during his visit. SK is supported by National Natural Science Foundation of China (NSFC) grant No. 12250610188 and Research Fund for International Scientists Project (W030231021012). XL is supported by NSFC grant No. 11501470, No. 11426187, No. 11791240561, the Fundamental Research Funds for the Central Universities 2682021ZTPY043 and partially supported by NSFC grant No. 11671328. Especially, XL would like to thank Bohui Chen, An-min Li, Guosong Zhao for their constant support and also thank all the friends met in different conferences. FY is supported by the NSFC grant No. 11950410490. RZ is supported by NSFC No. 12105198 and the High-level personnel project of Jiangsu Province (JSSCBS20210709).
\bigskip

\appendix

\section{Convention and notation}
We summarize the convention and notation used in this paper. Young diagrams or partitions are denoted by the Greek letter $\mu, \nu,\rho, \cdots$. A Young diagram $\lambda$ is given as $\lambda = \big((\lambda)_1, (\lambda)_2, \cdots \big) $, where $(\lambda)_{i}$ denotes the number of boxes in the $i$-th row of $\lambda$. A box at the $i$-th row and the $j$-th column is denoted as $(i,j)$. The transpose of a Young diagram $\lambda$ is denoted by $\lambda^T$. When multiple Young diagrams appear, we put lower indices to distinguish them as $\lambda_1, \lambda_2, \cdots $, which should be distinguished from $(\lambda)_1, (\lambda)_2, \cdots$ mentioned above. We use the vector symbol to denote these Young diagrams collectively, {\it i.e.,} $\vec{\lambda} = (\lambda_1, \lambda_2, \cdots )$. The total number of boxes of a given Young diagram $\lambda$ is denoted by $|\lambda|$, 
\begin{align}
|\lambda| = \sum_{(i,j)\in\lambda} 1 = \sum_{i} (\lambda)_i\ .
\end{align}
We also define a quantity for a given Young diagram $\lambda$ as
\begin{align}
\kappa(\lambda):=2\sum_{(i,j)\in\lambda}(j-i)\ ,
\end{align}
which corresponds to the second Casimir of the representation of the unitary group given by the Young diagram $\lambda$.

In the topological vertex formalism, we assign Young diagrams to each edge in the 5-brane web together with the arrow. Changing the direction of the arrow is equivalent to transposing the Young diagram as follows: 
\begin{center}
\begin{tikzpicture}
\draw (-3,0) -- (-1,0) ;
\draw[->,>=Latex] (-2,0) -- (-1.9,0) node [above] {$\lambda$};
\draw node (0,0) {$\mathbf{=}$} ;
\draw[->,>=Latex] (2,0) -- (1.9,0) node [above] {$\lambda^T$};
\draw (3,0) -- (1,0) ;
\end{tikzpicture}
\end{center}
We assign empty Young diagrams to the external lines that extend to infinity. 

Suppose we assign Young diagram $\lambda$ to the edge whose the corresponding K\"ahler parameter is given by $Q$, the contribution  from this edge is given by
\begin{align}\label{eq:edge-fac}
(-Q)^{|\lambda|} f_{\lambda}^{-(p_1 q_2 - p_2 q_1)}.
\end{align}
Here, $f_{\lambda}$ is the framing factor defined by
\begin{align}
f_{\lambda} = (-1)^{|\lambda|} q^{\frac12 \kappa(\lambda)}.
\end{align}
The power for the framing factor in \eqref{eq:edge-fac} is read off from the 5-brane charges $(p_1,q_1)$ and $(p_2,q_2)$ of the adjacent edges given in the following figure:
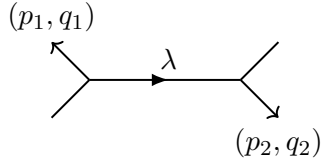
\begin{center}
\begin{tikzpicture}[scale=0.5]
\draw (-2,0) -- (2,0) ;
\draw [->](-2,0) -- (-3,1) node [above] {$(p_1,q_1)$};
\draw (-2,0) -- (-3,-1) ;
\draw  (2,0) -- (3,1) ;
\draw [->] (2,0) -- (3,-1) node [below] {$(p_2,q_2)$};
\draw [->,>=Latex] (0,0) -- (0.1,0) node [above] {$\lambda$};
\end{tikzpicture}
\end{center}

The contribution from each vertex, which is the intersection of the edges with Young diagram $\nu, \mu, \lambda$, respectively, is given by the topological vertex
\begin{align}
C_{\nu\mu\lambda}
& =q^{\frac{\kappa(\mu)}{2}+\frac{\kappa(\lambda)}{2}}s_{\lambda}(q^{-\rho})\sum_{\sigma}s_{\nu^T/\sigma}(q^{-\rho-\lambda})s_{\mu/\sigma}(q^{-\rho-\lambda^T})\ .
\label{topvertex-app}
\end{align}
As in the following figure, we read off the Young diagrams $\nu, \mu, \lambda$ clockwise:
\begin{center}
\begin{tikzpicture}[scale=0.5]
\draw (0,0) -- (-1,-1) ;
\draw[->,>=Latex] (-0.7,-0.7) -- (-0.71,-0.71) node [above left] {$\nu$};
\draw (0,0) -- (0,1.4) ;
\draw[->,>=Latex] (0,0.9) -- (0,0.91) node [above left] {$\mu$};
\draw (0,0) -- (1.4,0) ;
\draw[->,>=Latex] (0.9,0) -- (0.91,0) node [above right] {$\lambda$};
\node at (2.5,0) [right]  {$\mathbf{=}$} ;
\node at (4,0) [right]  {$C_{\nu\mu\lambda}$} ;
\end{tikzpicture}
\end{center}
Also, we have assumed that all the arrows from this vertex are outgoing. If the arrow is incoming, the corresponding Young diagram should be transposed.

\section{Identities of Nekrasov factor}\label{App:B}
The Nekrasov factor is defined by
\begin{equation}
N_{\lambda\nu}(Q;q):=\prod_{(i,j)\in \lambda}\lt(1-Qq^{1-i-j+(\lambda)_i+(\nu^T)_j}\rt)\prod_{(m,n)\in \nu}\lt(1-Qq^{m+n-(\lambda^T)_n-(\nu)_m-1}\rt),
\label{def:Nekra}
\end{equation}
which satisfies the following identities:
\begin{align}
N^{-1}_{\lambda\lambda}(1;q)
&=
(-1)^{|\lambda|}s_\lambda(q^{-\rho})s_{\lambda^T}(q^{-\rho})\ ,
\label{id:Schur-Nekra}
\\
N_{\lambda\nu}(Q;q)
&=
(-Q)^{|\lambda|+|\nu|}q^{\frac{1}{2}\kappa(\lambda)-\frac{1}{2}\kappa(\nu)} N_{\nu\lambda} \lt( Q^{-1};q \rt)\ ,
\label{Nekra-convert-formu}
\\
N_{\lambda\nu}(Q;q^{-1})
&= 
N_{\lambda^T\nu^T}(Q;q)\ ,
\\
N_{\lambda\nu}(Q;q)
&=
N_{\nu^T \lambda^T}(Q;q)\ ,
\\
N_{\varnothing\varnothing}(Q;q)
&=
1\ .
\label{eq:N-sym}
\end{align}

The Nekrasov factor is related to the factor defined by 
\begin{align}\label{eq:PE}
R_{\mu \nu} (Q;q) 
:= \prod_{i=1}^{\infty} \prod_{j=1}^{\infty}
\lt( 1 - Q q^{i+j-1-(\mu)_i-(\nu)_j} \rt)
\end{align}
as
\begin{align} \label{eq:R-definition}
R_{\mu \nu} (Q;q) 
= \mathrm{PE} \lt( - \frac{q}{(1-q)^2} Q \rt) N_{\mu^T \nu} (Q;q)\ ,
\end{align}
where PE is the plethystic exponential defined as
\begin{align}
\mathrm{PE}\big(f(x_i)\big) = \exp\bigg( \sum_{n=1}^{\infty} \frac1n f(x_i^n)\bigg)  \ . 
\end{align}
This factor appears in the following Cauchy identities:
\begin{align}
\sum_\lambda &~ Q^{|\lambda|} s_{\lambda/\mu}(q^{-\rho-\sigma})s_{\lambda/\nu}(q^{-\rho-\tau})
\cr
&~= R^{-1}_{\sigma \tau}(Q;q) \sum_\eta Q^{|\mu|+|\nu|-|\eta|}s_{\nu/\eta}(q^{-\rho-\sigma})s_{\mu/\eta}(q^{-\rho-\tau})\ ,\label{Schur-nor-id-spec}\\
\sum_\lambda &~ Q^{|\lambda|}s_{\lambda/\mu^T}(q^{-\rho-\sigma})s_{\lambda^T/\nu}(q^{-\rho-\tau})
\cr
&~=R_{\sigma \tau}(-Q;q)\sum_\eta Q^{|\mu|+|\nu|-|\eta|} s_{\nu^T/\eta}(q^{-\rho-\sigma})s_{\mu/\eta^T}(q^{-\rho-\tau})\ .
\label{Schur-twist-id-spec}
\end{align}

The new Nekrasov factor $\tilde{n}_{\lambda\nu}$, introduced in this paper, is related to the original Nekrasov factor $N_{\lambda\nu}$ as 
\begin{align}
\tilde{n}_{\lambda\nu}(a;\hbar) 
:= 
(- A)^{-\frac12 (|\lambda|+|\nu|)} q^{-\frac14 \hbar(\kappa(\lambda) - \kappa(\nu))}
N_{\lambda\nu}(A;q)\ ,
\end{align}
which is written explicitly as
\begin{align}
\tilde{n}_{\lambda\nu}(a;\hbar) 
= &\prod_{(i,j)\in \lambda}
2 \sinh \frac12 \Big(a + \hbar \big(1-i-j+(\lambda)_i+(\nu^T)_j \big)\Big)
\cr
& \quad \times
\prod_{(i,j)\in \nu}
2 \sinh \frac12  \Big(a + \hbar \big(i+j-1-(\lambda^T)_i-(\nu)_j \big)\Big)\ .
\end{align}
This satisfies the following identities:
\begin{align}
\tilde{n}_{\lambda\nu}(a;\hbar)
= \tilde{n}_{\nu^T \lambda^T}(a;\hbar)
= (-1)^{|\lambda|+|\nu|} \tilde{n}_{\nu\lambda}(-a;\hbar)
= (-1)^{|\lambda|+|\nu|} \tilde{n}_{\lambda^T \nu^T}(-a;\hbar)\ .
\end{align}
When one of the Young diagrams is empty, it is given by
\begin{align}
\tilde{n}_{\lambda^T \varnothing}(a;\hbar) 
= &\prod_{(i,j)\in \lambda^T}
2\, \sinh \frac12 \Big(a + \hbar (1-i-j+(\lambda^T)_i )\Big)
\cr
= &\prod_{(i,j)\in \lambda}
2\, \sinh \frac12\Big(a + \hbar (i-j )\Big)\ ,
\end{align}
which then leads to the following:
\begin{align}
\tilde{n}_{\lambda^T\varnothing}(a;\hbar)
= \tilde{n}_{\varnothing \lambda}(a;\hbar)
= (-1)^{|\lambda|} \tilde{n}_{\lambda \varnothing}(-a;\hbar)
= (-1)^{|\lambda|} \tilde{n}_{\varnothing\lambda^T}(-a;\hbar)\ .
\end{align}

\section{Proofs of the M-factor identity}\label{App:C}
In this section, we present two different proofs for the M-factor identity \eqref{eq:M-factor}. As given in \eqref{eq:M-factor-op-text}, it is defined as 
\begin{align}\label{eq:M-factor-op}
M_{\vec{\lambda}}(A_1,A_2,\cdots,A_N)
:= 
\frac{\bra{0}\mathbb{O}(q)\prod_{s=1}^N \Gamma_-(A_s q^{-\rho-\lambda_{s}})\ket{0}}{\bra{0}\mathbb{O}(q)\prod_{s=1}^N \Gamma_-(A_sq^{-\rho})\ket{0}}\ .
\end{align}
The square of the M-factor, $M^2_{\vec{\lambda}}
 = M_{\lambda_{1},\cdots,\lambda_{N}}\cdot M_{\lambda_{1},\cdots,\lambda_{N}}$, leads to the M-factor identity:
\begin{equation} \label{eq:rewriting}
 M^2_{\vec{\lambda}}(A_1,\cdots,A_N)
  =
 \prod_{s, t=1}^N  
 N^{-1}_{\lambda_{s}^T \lambda_t}(A_s A_t;q)
 \times \prod_{r=1}^N \prod_{f=1}^4 
 N_{\lambda_{r}^T \varnothing} (A_{r} \,{\sf M}_f^\ell ;q)\ , 
\end{equation}
where $\mathsf{M}_f$ are given in \eqref{eq:defM1234}, but for a short notation, we use $\mathsf{M}_f= \{\pm1, \pm q^\frac12\}$.

\paragraph{Proof 1:} 
It was noticed in \cite{Nawata:2021dlk} that the M-factor defined in \eqref{eq:M-factor-op} can be expressed as a factorized form of the off-diagonal and diagonal parts:
\begin{align}
M_{\vec{\lambda}}(A_1,A_2,\cdots,A_N)
= 
{}^{\rm o}\!M_{\vec{\lambda}}(A_1,A_2,\cdots,A_N) \times\prod_{n=1}^N {}^{\rm d}\!M_{\lambda_n}(A_n)\ ,
\label{def-M}
\end{align}
where the off-diagonal part takes the form
\begin{align}\label{eq:off-diag}
    &{}^{\rm o}\!M_{\vec{\lambda}}(A_1,A_2,\cdots,A_N)=\prod_{1\le s < t \le N} N^{-1}_{\lambda_s^T \lambda_t}(A_sA_t;q) \ .
\end{align}
and the diagonal part takes the form
\begin{equation}\label{eq:M-diag}
    {}^{\rm d}\!M_\lambda(A):=\prod_{(i,j)\in\lambda}\lt(1-A^{2}q^{i-j+(\lambda^T)_{j}-(\lambda)_{i}}\rt)\ .
\end{equation}

The square of the off-diagonal part is written as
\begin{align}\label{eq:off-diag-squared}
{}^{\rm o}\!{M}_{\vec{\lambda}}^2 
= \prod_{s=1}^N \prod_{t=1}^N N^{-1}_{\lambda_{s}^T \lambda_t}(A_s A_t;q)
\times 
\prod_{r=1}^N N_{\lambda_{r}^T \lambda_r}(A_r^2;q)\ .
\end{align}
As the first factor on the right-hand side of \eqref{eq:off-diag-squared} already appeared in \eqref{eq:rewriting}, it suffices to show that the diagonal part satisfies 
\begin{align}\label{eq:M-diag-id}
{}^{\rm d}\!M^2_\lambda(A){}= \
N^{-1}_{\lambda^T \lambda} (A^2;q) 
\prod_{f=1}^4 \prod_{\ell=\pm 1} N_{\lambda^T \varnothing} (A\, {\sf M}_f^{\ell} ;q)\ .
\end{align}
First, when $\lambda=\varnothing$, \eqref{eq:off-diag} yields 1 and the right-hand side of \eqref{eq:M-diag-id} also yields 1, and hence both sides of \eqref{eq:M-diag-id} are trivially the same. 
Next, when $\lambda\neq\varnothing$, rather than directly showing \eqref{eq:off-diag}, we show that both sides obey the same recursive relation when a box is added to the Young diagram $\lambda$. By induction, it then proves the identity \eqref{eq:M-diag-id}.

\begin{figure}
    \centering
\begin{tikzpicture}[x=0.75pt,y=0.75pt,yscale=-1,xscale=1]

\draw [line width=1.5]    (100,40) -- (342,41) ;
\draw [line width=1.5]    (100,40) -- (100,181) ;
\draw [line width=1.5]    (100,181) -- (150,181) ;
\draw [line width=1.5]    (150,122) -- (150,181) ;
\draw [line width=1.5]    (150,122) -- (221,122) ;
\draw [line width=1.5]    (221,90) -- (221,122) ;
\draw [line width=1.5]    (221,90) -- (271,90) ;
\draw [line width=1.5]    (271,57) -- (271,90) ;
\draw [line width=1.5]    (271,57) -- (342,57) ;
\draw [line width=1.5]    (342,41) -- (342,57) ;
\draw  [pattern=_32nusnocv,pattern size=6pt,pattern thickness=0.75pt,pattern radius=0pt, pattern color={rgb, 255:red, 0; green, 0; blue, 0}] (150,122) -- (166,122) -- (166,138) -- (150,138) -- cycle ;
\draw  [pattern=_j87jngvy3,pattern size=6pt,pattern thickness=0.75pt,pattern radius=0pt, pattern color={rgb, 255:red, 0; green, 0; blue, 0}] (100,181) -- (116,181) -- (116,197) -- (100,197) -- cycle ;
\draw  [pattern=_4fqllij1c,pattern size=6pt,pattern thickness=0.75pt,pattern radius=0pt, pattern color={rgb, 255:red, 0; green, 0; blue, 0}] (221,90) -- (237,90) -- (237,106) -- (221,106) -- cycle ;
\draw  [pattern=_0hz0ptd8u,pattern size=6pt,pattern thickness=0.75pt,pattern radius=0pt, pattern color={rgb, 255:red, 0; green, 0; blue, 0}] (271,57) -- (287,57) -- (287,73) -- (271,73) -- cycle ;
\draw  [pattern=_ngc26pfiu,pattern size=6pt,pattern thickness=0.75pt,pattern radius=0pt, pattern color={rgb, 255:red, 0; green, 0; blue, 0}] (342,41) -- (358,41) -- (358,57) -- (342,57) -- cycle ;
\draw  [fill={rgb, 255:red, 74; green, 144; blue, 226 }  ,fill opacity=1 ] (134,165) -- (150,165) -- (150,181) -- (134,181) -- cycle ;
\draw  [fill={rgb, 255:red, 74; green, 144; blue, 226 }  ,fill opacity=1 ] (205,106) -- (221,106) -- (221,122) -- (205,122) -- cycle ;
\draw  [fill={rgb, 255:red, 74; green, 144; blue, 226 }  ,fill opacity=1 ] (255,74) -- (271,74) -- (271,90) -- (255,90) -- cycle ;
\draw  [fill={rgb, 255:red, 74; green, 144; blue, 226 }  ,fill opacity=1 ] (326,41) -- (342,41) -- (342,57) -- (326,57) -- cycle ;
\draw  [fill={rgb, 255:red, 155; green, 155; blue, 155 }  ,fill opacity=1 ] (151,74) -- (167,74) -- (167,90) -- (151,90) -- cycle ;
\draw  [dash pattern={on 0.84pt off 2.51pt}]  (101,82) -- (159,82) ;
\draw  [dash pattern={on 0.84pt off 2.51pt}]  (159,41) -- (159,82) ;
\draw  [pattern=_u58tsbgr4,pattern size=6pt,pattern thickness=0.75pt,pattern radius=0pt, pattern color={rgb, 255:red, 19; green, 0; blue, 0}] (427,144) -- (443,144) -- (443,160) -- (427,160) -- cycle ;
\draw  [fill={rgb, 255:red, 74; green, 144; blue, 226 }  ,fill opacity=1 ] (428,170) -- (444,170) -- (444,186) -- (428,186) -- cycle ;

\draw (76,75) node [anchor=north west][inner sep=0.75pt]   [align=left] {$\displaystyle i$};
\draw (155,10) node [anchor=north west][inner sep=0.75pt]   [align=left] {$\displaystyle j$};
\draw (128,93) node [anchor=north west][inner sep=0.75pt]   [align=left] {$\displaystyle x=( i,j)$};
\draw (364,143) node [anchor=north west][inner sep=0.75pt]   [align=left] {$\displaystyle \mathfrak{A}( \lambda ) =\{$};
\draw (444,143) node [anchor=north west][inner sep=0.75pt]   [align=left] {$\displaystyle \}$};
\draw (364,169) node [anchor=north west][inner sep=0.75pt]   [align=left] {$\displaystyle \mathfrak{R}( \lambda ) =\{$};
\draw (445,169) node [anchor=north west][inner sep=0.75pt]   [align=left] {$\displaystyle \}$};
\end{tikzpicture}
    \caption{An example of a typical Young diagram $\lambda$. Each box (the gray one gives one representative) is labeled by the coordinate $x=(i,j)$, which indicates it is in the $i$-th row and $j$-th column. $\mathfrak{A}(\lambda)$ denotes the set of all boxes (with slashed pattern) that are allowed to be added to the Young diagram to give a new Young diagram. Similarly, $\mathfrak{R}(\lambda)$ denotes the set of all boxes (colored in blue) that can be removed from $\lambda$.}
    \label{fig:Young}
\end{figure}
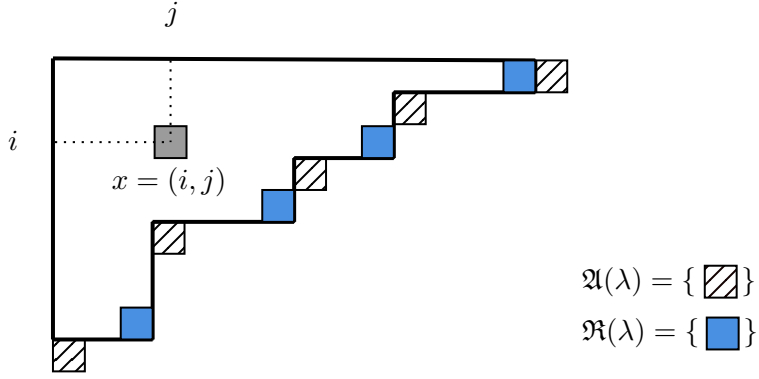

The recursive relations are as follows.  Let $x$ be a box in a Young diagram $\lambda$, then its position is expressed as $x=(i,j)$ and we introduce $\chi_x= q^{i-j}$. As shown in Eq. (D.18) in \cite{Nawata:2023wnk}, the diagonal part satisfies the following recursive relation, 
\begin{equation}\label{eq:recursion-dM-1}
\frac{{}^{\rm d}\!M_{\lambda+x}}{{}^{\rm d}\!M_{\lambda}} 
=
(1-A^2\chi_x^2)^2 \,\,\frac{\displaystyle\prod_{y\in \frakR(\lambda)}(1-A^2\chi_x\chi_y)}{\displaystyle\prod_{\substack{y\in \frakA(\lambda)}}(1-A^2\chi_x\chi_y)}\ ,
\end{equation}
where $\frakA(\lambda)$ denotes the set of boxes that can be added to $\lambda$, while $\frakR(\lambda)$ denotes the set of boxes that can be removed from $\lambda$, as illustrated in Figure \ref{fig:Young}.
\noindent For a compact notation, it is convenient to introduce 
\begin{align}\label{eq:def-of-Y-main}
Y_\lambda(Q;z)
:=\frac{\displaystyle\prod_{y\in \frakA(\lambda)}\Big(1-Q\frac{\chi_y}{z}\Big)}{\displaystyle\prod_{y\in \frakR(\lambda)}\Big(1-Q\frac{\chi_y}{z}\Big)}, \qquad 
Y_\varnothing(Q;z) = 1-\frac{Q}{z}\ .
\end{align}
We can then rewrite the square of \eqref{eq:recursion-dM-1} as 
\begin{equation}\label{eq:recursion-dM}
\frac{{}^{\rm d}\!M^2_{\lambda+x}}{{}^{\rm d}\!M^2_{\lambda}}
= (1-A^2\chi_x^2)^4\cdot Y^{-2}_{\lambda}(A^2;\chi_x^{-1})\ . 
\end{equation}

With $Y_\lambda(Q;z)$, it is straightforward to see that the Nekrasov factors satisfies the following recursion relations:
\begin{align}
&&\frac{N_{(\lambda_1+x)\lambda_2}(Q;q)}{N_{\lambda_1 \lambda_2}(Q;q)} 
&= Y_{\lambda_2}(Q; \chi_x)\ , \quad &
\frac{N_{\lambda_1 (\lambda_2+x)}(Q;q)}{N_{\lambda_1 \lambda_2}(Q;q)}
&= Y_{\lambda_1^T}(Q;\chi_x^{-1})\ ,&
\cr
&&\frac{N_{(\lambda_1+x)^T\lambda_2}(Q;q)}{N_{\lambda_1{}^T \lambda_2}(Q;q)} 
&= Y_{\lambda_2}(Q; \chi_x^{-1})\ ,\quad&
\frac{N_{\lambda_1 (\lambda_2+x)^T}(Q;q)}{N_{\lambda_1 \lambda_2^T}(Q;q)}
&= Y_{\lambda_1^T}(Q;\chi_x)\ , &
\label{Nekra-recursive-Y2-main}
\end{align}
which leads to 
\begin{equation}\label{eq:recursion-Y-main}
    \frac{Y_{\lambda+x}(Q;z)}{Y_{\lambda}(Q;z)}=\frac{(1-Q\chi_x q/z)(1-Q\chi_x q^{-1}/z)}{(1-Q\chi_x/z)^2}\ .
\end{equation}

We are now ready to show that the right-hand side of \eqref{eq:M-diag-id} also satisfies the identical relation as \eqref{eq:recursion-dM}.
To this end, we use \eqref{Nekra-recursive-Y2-main} and \eqref{eq:recursion-Y-main} to get 
\begin{align}\label{eq:recNxx}
\frac{N_{(\lambda+x)^T (\lambda+x)}(A^2;q)}{N_{\lambda^T\lambda}(A^2;q)}
&= \frac{N_{(\lambda+x)^T (\lambda+x)}(A^2;q)}{N_{(\lambda+x)^T\lambda}(A^2;q)}
 \frac{N_{(\lambda+x)^T \lambda}(A^2;q)}{N_{\lambda^T\lambda}(A^2;q)}
\cr
&= Y_{\lambda+x}(A^2;\chi_x^{-1}) Y_{\lambda}(A^2;\chi_x^{-1})
\cr
&= 
\frac{(1-A^2\chi^2_x q)(1-A^2\chi^2_x q^{-1})}{(1-A^2\chi^2_x)^2}
Y^2_{\lambda}(A^2;\chi_x^{-1})\ .
\end{align}
We also use \eqref{Nekra-recursive-Y2-main} to obtain
\begin{align}\label{eq:recNx}
\prod_{f=1}^4 \prod_{\ell=\pm 1} \frac{N_{(\lambda+x)^T \varnothing}(A\, {\sf M}_f^{\ell};q)}{N_{\lambda^T \varnothing}(A\, {\sf M}_f^{\ell};q)}
&= \prod_{f=1}^4 \prod_{\ell=\pm 1} (1 - A\, {\sf M}_f^{\ell} \chi_x)
\cr
&= (1-A^2\chi_x^2)^2 (1-A^2 q \chi_x^2) (1-A^2 q^{-1} \chi_x^2)\ ,
\end{align}
where $\mathsf{M}_f= \{\pm1, \pm q^\frac12\}$. It follows then from  \eqref{eq:recNxx} and \eqref{eq:recNx} that one finds that the right-hand side of \eqref{eq:M-diag-id} indeed satisfies the same recursive relation as \eqref{eq:recursion-dM}. 
Since the identity \eqref{eq:M-diag-id} is shown, \eqref{def-M} has been proven.\hfill     {\it Q.E.D.}

\paragraph{Proof 2:} \label{App:derivation}
Consider the definition of the M-factor given in \eqref{eq:M-factor-op}, which has the form of a normalized expectation value. Using \eqref{Overtex} and \eqref{eq:Gamma-pm}, we express the numerator as an operators expectation value:
\begin{align}\label{eq:M-XY}
\bra{0}\mathbb{O}(q)\prod_{s=1}^N\Gamma_-( A_s q^{-\rho-\lambda_{s}})\ket{0}
= \bra{0} e^X e^Y \ket{0},
\end{align}
where
\begin{align}\label{eq:defofXY}
X= & \sum_{n=1}^{\infty} \lt( - \frac{1}{2n} \frac{1+q^{n}}{1-q^{n}} J_{2 n} + \frac{1}{2n} J_{n} J_{n} \rt)  \ , 
\cr
Y= & \sum_{s=1}^N \sum_{n=1}^\infty \sum_{i=1}^{\infty} \frac{1}{n} 
\lt( A_s q^{i-\frac12-(\lambda_{s})_i} \rt) ^n J_{- n}\ ,
\end{align}
where the operators $J_n$ satisfy \eqref{eq:Commutation-J}. Note that the denominator of the M-factor 
in \eqref{eq:M-factor-op} is obtained by taking $\lambda_s\!=\!\varnothing$. So, the M-factor identity \eqref{eq:rewriting} can be re-expressed~as  
\begin{align}\label{eq:M2asXY}
    M^2_{\vec{\lambda}}(A_1,\cdots,A_N) = \Bigg(\frac{\bra{0} e^X e^Y \ket{0} }{\bra{0} e^X e^Y \ket{0}\Big|_{\vec{\lambda} = \vec{\varnothing}}} \Bigg)^2\ . 
\end{align}

It follows from the Baker-Campbell-Hausdorff formula, given in an alternative form~\cite{Hayashi:2020hhb}, 
\begin{align}
e^X e^Y = e^Y e^{X+[X\,,\,Y]+\frac12 \lt[[X\,,\,Y]\,,\,Y\rt] + \frac{1}{3!} \lt[\lt[[X\,,\,Y]\,,\,Y\rt],\,Y\rt] + \cdots}\ ,
\end{align}
and $\bra{0} e^Y \!= \!\bra{0}$ due to $\bra{0} J_{-n}\! =\! 0$ for $ n \ge 1$, that the expectation value \eqref{eq:M-XY} is rewritten~as
 \begin{align}\label{eq:0XY0}
\bra{0} e^X e^Y \ket{0} 
= \bra{0} e^{X+[X\,,\,Y]+\frac{1}{2!} [[X\,,\,Y]\,,\,Y] + \frac{1}{3!} [[[X\,,\,Y]\,,\,Y]\,,\,Y] + \cdots} \ket{0}\ . 
\end{align}

We then compute each commutator term in the exponent. We first evaluate the commutator between $X$ and $Y$ as
\begin{align}\label{eq:XY}
&[X,Y] 
\cr
&= \lt[ \sum_{n=1}^{\infty} 
\lt( - \frac{1}{2n} \frac{1+q^{n}}{1-q^{n}} J_{2 n} + \frac{1}{2n} J_{n} J_{n}\rt)\,  {}_{\textstyle,} \, 
\sum_{s=1}^N \sum_{m=1}^\infty \sum_{i=1}^{\infty} \frac{1}{m}
\lt( A_s q^{i-\frac12-(\lambda_{s})_i} \rt) ^m J_{- m}\rt]
\cr
&=  
\sum_{s=1}^N \sum_{i=1}^{\infty} 
\sum_{m,n=1}^{\infty} 
\biggl( - \frac{1}{2n} \frac{1+q^{n}}{1-q^{n}}  
\frac{1}{m} \lt( A_s q^{i-\frac12-(\lambda_{s})_i} \rt) ^m 
2n \, \delta_{2n, m}
\cr
& \qquad \qquad \qquad \qquad \qquad 
+ \frac{1}{2nm} \lt( A_s q^{i-\frac12-(\lambda_{s})_i} \rt) ^m 
2 n \, \delta_{n,m} J_n \biggr)
\cr
&=  
\sum_{s=1}^N \sum_{i=1}^{\infty} \sum_{n=1}^{\infty} 
\lt(
- \frac{1}{2 n} \frac{1+q^{n}}{1-q^{n}}  
\lt( A_s q^{i-\frac12-(\lambda_{s})_i} \rt) ^{2n} 
+ \frac{1}{n}  
\lt(A_s q^{i-\frac12-(\lambda_{s})_i} \rt) ^n J_n
\rt) ,
\end{align}
where in the second equality, we used $[J_{2n},J_{-m}]=2n \delta_{2n,m}$ and $[J_n J_n,J_{-m}]= 2n \delta_{n,m} J_n$, which follow from the commutation relation \eqref{eq:Commutation-J}. 

Next, the commutator of \eqref{eq:XY} with $Y$ becomes
\begin{align}\label{eq:XYY}
&\big[[X,Y],Y\big] 
\cr
&=  
\Bigg[ 
\sum_{s=1}^N \sum_{i=1}^{\infty} \sum_{n=1}^{\infty} 
\lt(
- \frac{1}{2 n} \frac{1+q^{n}}{1-q^{n}} 
\lt( A_s q^{i-\frac12-(\lambda_{s})_i} \rt) ^{2n} 
+ \frac{1}{n}  
\lt(A_s q^{i-\frac12-(\lambda_{s})_i} \rt) ^n J_n
\rt) 
\, {}_{\textstyle,} 
\cr
& \qquad 
\sum_{t=1}^N \sum_{j=1}^\infty \sum_{m=1}^{\infty} \frac{1}{m}
\lt( A_t q^{j-\frac12-(\lambda_{t})_j} \rt) ^m J_{- m}
\Bigg]
\cr
&= \sum_{s,t=1}^N \sum_{i,j =1}^{\infty} \sum_{m,n=1}^{\infty} 
\frac{1}{n}  
\lt(A_s q^{i-\frac12-(\lambda_{s})_i} \rt) ^n 
\frac{1}{m} \lt( A_t q^{j-\frac12-(\lambda_{t})_j} \rt) ^m
n\, \delta_{n,m}
\cr
&= \sum_{s,t=1}^N 
\sum_{i,j=1}^{\infty} 
\sum_{n=1}^{\infty}   
\frac{1}{n}  
\lt( A_s A_t q^{i + j - 1 - (\lambda_{s})_i -(\lambda_{t})_j}  \rt)^n,
\end{align}
where we used $[J_{n},J_{-m}]= n \delta_{n,m}$ in the second equality. Notice that $\big[[X,Y],Y\big]$ does not involve $J_{\pm n}$ any more. This means that all the higher order commutators appearing in \eqref{eq:0XY0} vanish:
\begin{align}\label{eq:XYYY}
\big[\big[[X,Y],Y\big],Y\big] = \big[\big[\big[[X,Y],Y\big],Y\big],Y\big] = \cdots = 0\ .
\end{align}

Now, putting all the nontrivial commutators into  \eqref{eq:0XY0}, one finds that the result expectation value is the exponential of a polynomial of the operators $J_{n}$ with $n \ge 1$. As $J_n \ket{0} = 0$ for $n \ge 1$, any polynomial of $J_{n\ge1}$ does not contribute. The remaining non-vanishing terms then amount to  
\begin{align}\label{eq:eXeY-1}
\bra{0} e^X e^Y \ket{0} 
& = \prod_{s=1}^N \prod_{i=1}^{\infty} 
\exp \lt( - \frac12 \sum_{n=1}^{\infty}
\frac{1}{n} \frac{1+q^{n}}{1-q^{n}}    
\lt( A_s q^{i-\frac12-(\lambda_{s})_i} \rt) ^{2n} 
\rt)
\cr
& \qquad 
\times 
\prod_{s, t=1}^N 
\prod_{i, j=1}^{\infty} 
\exp \left( 
\frac12 \sum_{n=1}^{\infty}   
\frac{1}{n}  
\lt( A_s A_t q^{i + j - 1 - (\lambda_{s})_i -(\lambda_{t})_j}  \rt) ^n
\right),
\end{align}
where $\bra{0} \! 0 \rangle = 1$. In order to perform the summation over the index $n$ in this expression, we first rewrite the factor in the exponent in the first line to the following form
\begin{align}\label{eq:trick}
\frac{1+q^n}{1-q^n} 
&=  \frac{(1+q^n)^2}{1-q^{2n}}
= (1 + 2q^n + q^{2n}) \sum_{j=1}^{\infty} q^{2n(j-1)}
\cr
&=  
\sum_{j=1}^{\infty} \lt( ( q^{j-1} )^{2n}+ 2 ( q^{j-\frac12} )^{2n}+ ( q^{j} )^{2n} \rt).
\end{align}
Then, denoting $\mathsf{x} := A_s q^{i-\frac12-(\lambda_{s})_i}$, it follows from
\eqref{eq:trick} that we perform the summation over $n$ in the first line of \eqref{eq:eXeY-1} as follows: 
\begin{align}
\sum_{n=1}^{\infty} \frac{1}{n} \frac{1+q^n}{1-q^n} \mathsf{x}^{2n}
&= \sum_{j=1}^{\infty} \sum_{n=1}^{\infty}  \frac{1}{n}
\lt( ( q^{j-1} \mathsf{x})^{2n}+ 2 ( q^{j-\frac12} \mathsf{x})^{2n}+ ( q^{j} \mathsf{x})^{2n} \rt)
\cr
&= \sum_{j=1}^{\infty} \sum_{f=1}^4 \sum_{\ell= \pm 1} \sum_{n=1}^{\infty} \frac{1}{n} ( {\sf M}_f^{\ell} q^{j-\frac12} \mathsf{x})^{n}
\cr
&= -\sum_{j=1}^{\infty} \sum_{f=1}^4 \sum_{\ell= \pm 1} \log \lt( 1 - {\sf M}_f^{\ell} q^{j-\frac12} \mathsf{x} \rt),
\end{align}
where $\mathsf{M}_f= \{\pm1, \pm q^\frac12\}$.
This leads to
\begin{align}
\exp \lt( - \frac12 \sum_{n=1}^{\infty} \frac{1}{n} \frac{1+q^n}{1-q^n} \mathsf{x}^{2n} \rt)
=& \prod_{j=1}^{\infty} \prod_{f=1}^4 \prod_{\ell= \pm 1} \lt( 1 - {\sf M}_f^{\ell} q^{j-\frac12} \mathsf{x} \rt)^{\frac12}.
\end{align}
Likewise, by denoting $\mathsf{y} := A_s A_t q^{i+j-1-(\lambda_{s})_i-(\lambda_{t})_j}$, the second line of \eqref{eq:eXeY-1} reads,
\begin{align}
\exp\lt( \frac12 \sum_{n=1}^{\infty} \frac{1}{n}\mathsf{y}^n \rt) 
= \exp \lt( - \frac12 \log (1-\mathsf{y}) \rt)
= (1-\mathsf{y})^{-\frac12}.
\end{align} 

Putting all together, we find
\begin{align} \label{eq:XY2}
\Big(\bra{0} e^X e^Y \ket{0} \Big)^2
&=  \prod_{r=1}^N \prod_{f=1}^4 \prod_{\ell= \pm 1} \prod_{i, j=1}^{\infty} \lt( 1 - {\sf M}_f^{\ell} A_r q^{i+j-1-(\lambda_{r})_i} \rt)
\cr
& \quad 
\times 
\prod_{s,t=1}^N \prod_{i, j=1}^{\infty} \lt( 1 - A_s A_t q^{i + j - 1 - (\lambda_{s})_i -(\lambda_{t})_j}\rt)^{-1}
\cr
&=  \prod_{r=1}^N \prod_{f=1}^4 \prod_{\ell= \pm 1} 
R_{\lambda_{r} \varnothing} ({\sf M}_f^{\ell} A_r) 
\times 
\prod_{s,t=1}^N 
R^{-1}_{  \lambda_{s} \lambda_{t} } (A_s A_t) .
\end{align}
We use \eqref{eq:R-definition} to rewrite \eqref{eq:XY2} in terms of the Nekrasov factors:
\begin{align}\label{eq:PEN}
\Big(\bra{0} e^X e^Y \ket{0}\Big)^2
&= \prod_{r=1}^N \prod_{f=1}^4 \prod_{\ell= \pm 1}\mathrm{PE}\Big(-\frac{q}{(1-q)^2}{\sf M}_f^{\ell} A_r\Big) 
N_{\lambda^T_{r} \varnothing} ({\sf M}_f^{\ell} A_r)\cr
&\qquad\times 
\prod_{s,t=1}^N \mathrm{PE} \Big(\frac{q}{(1-q)^2}A_s A_t\Big) 
N^{-1}_{ \lambda^T_{s} \lambda_{t} } (A_s A_t) \ .
\end{align}
As the Nekrasov factor associated with the empty Young diagrams is trivial $N_{\varnothing\varnothing}(Q)=1$, it readily reproduces the perturbative part: 
\begin{align}
\Big(\left. \bra{0} e^X e^Y \ket{0} \right|_{\vec{\lambda}=\vec{\varnothing}}\Big)^2 &=
\mathrm{PE}\bigg( 
\frac{q}{(1-q)^2}\Big( -\sum_{r=1}^{N}\sum_{f=1}^4\sum_{\ell=\pm} \mathsf{M}^\ell_fA_r
+\sum_{s,t=1}^{N}A_sA_t
\Big)
\bigg)\cr
&=\mathrm{PE} \Big(
\frac{2q}{(1-q)^2} \sum_{s<t}
A_s A_t
\Big)\ .
\end{align}
Finally, plugging the above into \eqref{eq:M2asXY},
one arrives at \eqref{eq:rewriting}:
\begin{align}
\frac{\Big(\bra{0} e^X e^Y \ket{0}\Big)^2}{\Big(\bra{0} e^X e^Y \ket{0}\Big|_{\vec{\lambda} = \vec{\varnothing}}\Big)^2} 
& 
= \prod_{r=1}^N \prod_{f=1}^4 \prod_{\ell= \pm 1} 
N_{\lambda_{r}^T \varnothing} ({\sf M}_f^{\ell} A_r)
\times 
\prod_{s=1}^N \prod_{t=1}^N 
N^{-1}_{\lambda_{s}^T \lambda_{t} } (A_s A_t) \qquad
\hfill{\it Q.E.D.}  \nonumber
\end{align} 

\bigskip
\bibliographystyle{JHEP}
\bibliography{ref}
\end{document}